\documentstyle[12pt]{article}
\newcommand{\slq}{\raise.15ex\hbox{$/$}\kern-.57em\hbox{$q$}}
\newcommand{\slp}{\raise.15ex\hbox{$/$}\kern-.57em\hbox{$p$}}
\newcommand{\be}{\begin{equation}}
\newcommand{\ee}{\end{equation}}
\newcommand{\bear}{\begin{eqnarray}}
\newcommand{\ear}{\end{eqnarray}}

\newcommand{\Tr}{{\rm{Tr}}}
\newcommand{\M}{{\cal{M}}}
\newcommand{\gsim}{\stackrel{\scriptstyle >}{\sim}}
\newcommand{\lsim}{\stackrel{\scriptstyle <}{\sim}}
\date{}
\renewcommand{\theequation}{\arabic{section}.\arabic{equation}}

\begin{document}
\begin{titlepage}
\begin{flushright}
HD--THEP--95--47
\end{flushright}
\quad\\
\vspace{1.8cm}
%Korrigierte Version vom 20. 3. 1996
\begin{center}
{\bf\LARGE Integrating out Gluons}\\
\medskip
{\bf\LARGE in Flow Equations}\\
\vspace{1cm}
Christof Wetterich\\
\bigskip
Institut  f\"ur Theoretische Physik\\
Universit\"at Heidelberg\\
Philosophenweg 16, D-69120 Heidelberg\\
\vspace{1cm}
{\bf Abstract}
\end{center}
We present an exact nonperturbative flow equation for the average
action for quarks which incorporates the effects of gluon
fluctuations. With suitable truncations this allows one to
compute effective multiquark interactions in dependence on an
infrared scale $k$. Our method amounts to integrating out the gluons
with momenta larger than $k$.
\end{titlepage}
\newpage
\section{Introduction}
Effective multiquark interactions have often served
as a starting point for QCD-motivated investigations of hadronic
physics. A well-known example is the Nambu-Jona-Lasinio model
\cite{1} which is based on an effective four-quark interaction. The
idea is that these models are valid for quark fluctuations with momenta
smaller than a given scale $k$ whereas fluctuations with higher
momenta and gluons are thought  to be already integrated out. In
the more fundamental context of QCD one would like to have some
information on the strength of such an interaction in dependence
on $k$. Furthermore, in a realistic theory the effective four-quark
interactions will be momentum-dependent. For heavy quarks this
momentum dependence translates into the form of the
heavy quark potential which is the basis for the non-relativistic
quark model for charm and bottom-bound states \cite{2}.
Recently a QCD-motivated effective four-quark interaction (at
a scale $k\approx 1.5$ GeV) has also been used to derive the emergence
of mesonic bound states and chiral symmetry breaking from the
solution of nonperturbative flow equations \cite{Ell}. In
order to make direct contact with QCD all these approaches
need a method capable of integrating out the gluonic degrees
of freedom.

The problem of integrating out completely the gluons seems almost
as difficult as the full solution of QCD. Even if an effective
infrared cutoff is present in the quark sector the gluons will
lead to strong nonlocalities in the quark interactions. Formally, one
would have to derive the effective action for gluons and quarks, if
needed with an additional infrared cutoff $\sim k$ in the quark sector.
{}From there the gluonic field equations can be derived, giving the
(classical) gluon field $A_\mu^{(cl)}$ as a functional of the quark
field $\psi$. Inserting this solution into the effective action then
leads to an effective action for the quarks from which the effective
multiquark interactions can be extracted. Obviously such a program
must become very complicated since due to confinement a purely
gluonic description of the gauge degrees of freedom seems not very
appropriate in the low momentum range. New non-perturbative effective
degrees of freedom are expected to become relevant for very low
momenta below the confinement scale $\Lambda$.

For many practical purposes, however, the extreme low momentum
behaviour of the effective quark interactions is actually not
needed. For example, the use of the effective heavy quark potential
for the low lying charmonium and bottonium states essentially needs
information about the effective four-quark interaction at momenta
$q^2\gsim(300\ {\rm MeV})^2$. Since the quark momenta act as
an effective infrared cutoff also for the gluon
fluctuations\footnote{We work here in the Euclidean regime
where $q^2>0$ and quarks are off-shell.},
we do not expect that confinement effects are dominant in the
range of sufficiently large momenta. One may therefore
think of using there an approximation
with an additional infrared cutoff $k$ both for the gluon and the
quark degrees of freedom. This idea is realized by the concept of the
average action $\Gamma_k$ \cite{W1} which obtains from integrating out
all quantum fluctuations with momenta larger than $k$. For a given
$k>0$ one does not expect that $\Gamma_k$ always describes very
accurately the $n$-point functions for momenta smaller than $k$ -
in particular the confinement physics in QCD is cut out for large
enough $k$. On the other hand the effective
average action can be interpreted
as a good approximation to the generating functional for 1PI Green
functions with momenta sufficiently larger than $k$.
A possible first approach for a computation of the potential in the
range relevant for heavy quarkonia could therefore be
attempted by a computation of the quark average action $\Gamma_k[\psi]$
for $k$ somewhat larger than the confinement scale. We will see next
how this can be improved by lowering $k\to0$, provided the momenta
in the effective quark $n$-point functions are sufficiently large.

The dependence of the average action on the scale $k$ is described
by an exact non-perturbative flow equation \cite{4E}. This functional
differential equation needs a truncation in order to become approximately
solvable in a non-perturbative context. Typically, one may retain only
gluonic two-, three- and four-point functions in the gauge sector.
The question arises in what context this type of truncation can lead to
meaningful results. First we should point out that such a truncation
is not restricted to the range of applicability of perturbation theory.
It is general enough to account for a nonperturbative form of the
gluon propagator and vertices. Nevertheless, a simple gluonic
description is not expected to remain valid once effective
non-perturbative degrees of freedom different from the gluons play
an important role. (This is partly related to situations where higher
$n$-point functions for the gluons become important.) The range
in $k$ where a ``gluonic truncation'' of the average action
can be trusted is limited
by this fact. For $k$  larger than  the confinement
scale the precise description of the confinement mechanism should not
be of dominant importance. For $k$ as large as a few GeV one
expects that a perturbative description of $\Gamma_k$ becomes valid.
It is well conceivable that
once $k$ is lowered, there first appears
a region  where perturbation theory breaks down
but a truncation to a few gluon Green functions still remains a valid
approximation. Let us now discuss what happens if one tries to
follow the flow of the average action beyond this region
towards $k\to0$ within such a ``gluonic truncation''.
One expects that the low momentum behaviour of the $n$-point functions
becomes only poorly described and may even give a qualitatively
wrong picture. In contrast, the Green functions with momenta
$q^2\gsim (300\ {\rm MeV})^2$ should only mildly be affected by the
error from the insufficient truncation. Since an infrared cutoff is
already present due to the nonvanishing momenta the variation of
an additional and smaller infrared cutoff $k$ should not be very
relevant anymore. We conclude that a ``gluon truncation'' can give
reliable results even for $k\to0$ (all fluctuations included), provided
the momenta of the Green functions are sufficiently large. In particular,
one wonders if a computation of the heavy quark potential $V(R)$ for the
range in $R$ relevant for the low-lying bound states may be
feasible along these lines.

Let us next turn to the light mesons. The scale for the
formation of mesonic bound states was found \cite{Ell} to be around
$k_\varphi\approx $ 700 MeV. For $k<k_\varphi$ a mesonic description
becomes appropriate \cite{Ell}. Following the
flow equation of an effective quark-meson model \cite{Jun} for
$k<k_\varphi$ one finds that chiral symmetry breaking
sets in at $k\approx$ 450 MeV.
Subsequently the quarks acquire a constituent mass $m_q\approx$ 350 MeV
from their Yukawa coupling to the chiral symmetry-breaking
order parameter. This implies that quarks effectively decouple for
$k\lsim$ 300 MeV. Since mesons do not carry
colour one may hope that the details of the
confinement mechanism are not crucial for an understanding of
the meson mass spectrum and decays. This may hold despite the
fact that already the formation of light meson bound states is
certainly a highly non-perturbative effect. The exchange of
many gluons needs to be considered for the description of
bound states.

In order to understand this issue more precisely one may
imagine for a moment that at some high scale $k_i$ the classical
gluon exchange gives rise to an ``initial'' effective four-quark
interaction $\sim\bar\psi\gamma^\mu\psi\frac{g^2}{q^2}\bar\psi\gamma_
\mu\psi$ and the gluons are neglected for $k<k_i$. If one follows
the flow equations for the fermionic model for $k<k_i$ the presence
of an ``initial'' four-fermion coupling influences the trajectories
and results for low enough $k$ in a complicated structure of the
effective four-quark interaction. In this sense a great part
of ``multi-gluon exchanges'' is already incorporated in the
solution of the purely fermionic flow equation. Nonperturbative
gluon effects arise even though the gluons do not appear
explicitly for $k<k_i$. This is, however, not the whole story:
There are also contributions from gluon fluctuations which are
not included in the standard flow of an effective fermionic theory.
Contributions from those ``residual gluon fluctuations''
lead to important
corrections to the standard fermionic flow equations. They
will be computed explicitly in the present paper.
Corrections to the flow of the four-quark interaction involve the
two-, three- and four-point functions for gluons.
In this context the truncation in
the gluon sector only enters through the ``residual gluon
contributions''. In an appropriate formulation the ``residual
gluons'' influence only very indirectly the mesons
via their effects on the quarks. For $k<300$ MeV
their contributions are suppressed by the quark constituent masses.
It is in this type of formulation that the precise description of
confinement may not be needed, and even a relatively inaccurate
``gluon truncation'' may not destroy the reliability of a calculation
of meson properties for $k\to0$.

In order to address the questions raised in this introduction more
quantitatively one needs a method
which  integrates out the gluon fluctuations
with momenta $q^2>k^2$. Formally, it is
easy to do so at a given scale $k_1$. One needs to compute the
effective average
action for quarks and gluons $\Gamma_{k_1}[\psi,A]$ at this scale.
Solving the
classical field equation for the gluon field $A$ in dependence on
$\psi$ and
inserting this classical solution into $\Gamma_{k1}[\psi,A]$ yields
exactly an effective average action
$\Gamma_{k_1}[\psi]$ involving
only the quark fields. The quark-effective action $\Gamma_{k_1}[\psi]$
may then be used as an initial value for solving a pure fermionic flow
equation for $k<k_1$. Obviously, the shortcoming of such an approach
ist the complete omission of the effects of residual
gluon fluctuations with
$q^2<k_1^2$. Choosing a different scale $k_2$ for eliminating the gluons
will lead to a different result. Such a sharp transition between the
quark-gluon system and a description involving only quarks necessarily
introduces a certain degree of arbritariness.

We propose here a more refined method which changes the classical
solution for the gluon field in the course of the evolution towards
lower $k$, thus reflecting the change in the form of $\Gamma_k[\psi,A]$.
The result is a smooth procedure for integrating out the gluons,
where at every scale $k$ all gluon fluctuations with $q^2>k^2$ are
included. Nevertheless, we obtain a flow equation for the quark
effective
average action $\Gamma_k[\psi]$, where gluon fields do not appear
explicitly. The inclusion of contributions from
residual gluon fluctuations induces correction
terms in the flow equation for the quark interactions. In
particular, we choose here a formulation where the $k$-dependent
classical solution for $A$ as a functional of $\psi$ includes an
effective infrared cutoff $\sim k$. As a consequence, the only
nonlocalities
in $\Gamma_k[\psi]$ concern length scales shorter than $k^{-1}$. For
the fermionic low momentum modes $\Gamma_k[\psi]$ is an
effectively local action. For example, a derivative expansion is
meaningful for $q^2\ll k^2$. The expected nonlocalities arising from
the complete elimination of gluon fields (for example a four-quark
interaction $\sim\frac{1}{q^2}$ appearing already in the Born
approximation) build up only step by step as $k$ is lowered to
zero. As a result of this method we will end with an exact
nonperturbative
flow equation for the scale dependence of $\Gamma_k[\psi]$.
Approximations will be needed to solve this equation but they are
not limited to perturbative concepts. For example, we may neglect
$\psi^6$ interactions which should be roughly equivalent to
neglecting the influence of the baryons on the meson properties.
Also the correction terms reflecting the residual
gluon fluctuations require limited knowledge about the effective gluon
propagator and vertices. It is hoped that rather crude approximations
for the gluonic vertices can already lead to satisfactory results.

In sect. 2 we develop our formalism for a simple model of two
scalar fields. The flow of the effective average action for one of
the scalar fields obtains by integrating out the other scalar field
at any scale $k$. This is generalized to quarks and gluons in sect. 3.
In sect. 4 we give a first demonstration how this formalism describes
the flow of the two- and four-point function in the effective quark
theory. The special case of heavy quarks is addressed in sect. 5. Here
we argue that only the evolution of the gauge field propagator is needed
in this limit and we compute the corresponding flow equation in
one of the appendices.
Finally, our conclusions
are contained in sect. 6.

\section{Reduction of degrees of freedom}
\setcounter{equation}{0}
In this section we consider for simplicity two types of scalar fields,
$\varphi$ and $\psi$. We want to develop a formalism how to translate
evolution equations for the effective average action for $\varphi$
and $\psi$
into corresponding equations involving only $\psi$. The reader may
associate $\varphi$ with the gluon fields and $\psi$ with the quark
fields.
Our aim is then the construction of the effective average action for
quarks
out of the coupled quark-gluon system. This amounts to integrating
out the
gluonic degrees of freedom represented in the simplified model of this
section by $\varphi$. We start with the scale-dependent generating
functional
for the connected Green functions
\be\label{2.1}
W_k[J,K]=\ln\int D\varphi'D\psi'\exp-
\bigl\lbrace S[\varphi',\psi']
+\Delta^{(\varphi)}_kS[\varphi']+\Delta_k^{(\psi)}S[\psi']-J^\dagger
\varphi'-K^\dagger\psi'\bigr\rbrace\ee
Here we denote the degrees of freedom contained in $\varphi'$
(for example
the Fourier modes) by ${\varphi'}^\alpha$
and similar for $\psi'$, $J$ and $K$, with\footnote{We use indices
$\alpha,\alpha'$ etc. for $\varphi$ and $\beta,\beta'$ etc. for $\psi$.}
\be\label{2.2}
J^\dagger\varphi'=J^*_\alpha{\varphi'}^\alpha,\quad K^\dagger\psi'=
K_\beta^*{\psi'}^\beta\ee
We have introduced an infrared cutoff quadratic in the fields
\be\label{2.3}
\Delta^{(\varphi)}_kS[\varphi']=\frac{1}{2}{\varphi'}^\dagger
R^{(\varphi)}_k\varphi'\ee
and similar for $\psi$. This suppresses the contribution of fluctuations
with small momenta $q^2<k^2$ to the functional integral (\ref{2.1}).
Typically
$R^{(\varphi)}_k,R_k^{(\psi)}$ are functions of $q^2$ with the properties
\be\label{2.4}
R_k\sim\left\lbrace\begin{array}{lll}
k^2& {\rm for}& q^2\ll k^2\\ q^2\exp(-\frac{q^2}{k^2})&{\rm for}&
q^2\gg k^2
\end{array}\right.\ee
as, for example,
\be\label{2.5}
R_k^{(\varphi)}=\frac{Z_kq^2\exp\left(-\frac{q^2}{k^2}\right)}{1-\exp
\left(-\frac{q^2}{k^2}\right)}\ee
The effective average action $\Gamma_k[\varphi,\psi]$ is related to
the Legendre transform of $W_k[J,K]$
\be\label{2.6}
\tilde\Gamma_k[\varphi,\psi]=-W_k[J,K]+J^\dagger\varphi+K^\dagger\psi\ee
\be\label{2.7}
\varphi^\alpha=\frac{\partial W_k}{\partial J^*_\alpha},\quad \psi^\beta
=\frac{\partial W_k}{\partial K^*_\beta}\ee
by subtracting the infrared cutoff term
\be\label{2.8}
\Gamma_k[\varphi,\psi]=\tilde\Gamma_k[\varphi,\psi]-\Delta_k^{(\varphi)}S
[\varphi]-\Delta_k^{(\psi)}S[\psi]\ee
For $k\to0$ the infrared cutoff $\Delta_kS=\Delta_k^{(\varphi)}S+\Delta_k
^{(\psi)}S$ vanishes and $\Gamma_0$ is the usual generating function for
the 1PI Green functions. Using the quadratic form of $\Delta_kS$ it is
straightforward to derive an exact non-perturbative
evolution equation for the dependence
of the effective average action on the scale $k$ $(t=\ln k)$ \cite{4E}
\be\label{2.9}
\frac{\partial\Gamma_k}{\partial t}=\frac{1}{2}Tr\left\lbrace
(\tilde\Gamma_k^{(2)})^{-1}\frac{\partial R_k}{\partial t}
\right\rbrace\ee
Here $\tilde\Gamma_k^{(2)}=\Gamma_k^{(2)}+R_k$ and the inverse propagator
$\Gamma_k^{(2)}$ is the second functional derivative of $\Gamma_k$ with
respect to the fields. The matrix
$R_k=R_k^{(\varphi)}+R_k^{(\psi)}$ is block diagonal in $\varphi$ and
$\psi$
spaces.
The presence of the infrared cutoff $R_k$ in
$\tilde \Gamma_k^{(2)}$ guarantees infrared finiteness for the momentum
integral implied by the trace even in case of massless modes. Ultraviolet
finiteness is guaranteed by the exponential decay of $\partial
R_k/\partial
t$ (\ref{2.5}). A solution of the flow equation (\ref{2.9}) interpolates
between the classical action for $k\to\infty$ (or $k$ equal to
some ultraviolet cutoff $\Lambda$) and the effective action for $k\to0$.

The generating functional for the connected Green functions for $\psi$
obtains
from (\ref{2.1}) for $J=0$
\be\label{2.10}
W_k[K]\equiv W_k[J=0,K]\ee
Correspondingly, we may
introduce an effective action expressed only in terms of
$\psi$
\be\label{2.11}
\tilde\Gamma_k[\psi]=\tilde\Gamma_k[\varphi_k[\psi],\psi]\ee
\bear\label{2.12}
\Gamma_k[\psi]&=&\tilde\Gamma_k[\psi]-\Delta_k^
{(\psi)}S[\psi]\nonumber\\
&=&\Gamma_k[\varphi_k[\psi],\psi]+\Delta_k^{(\varphi)}S[\varphi_k[\psi]]
\ear
by inserting the $k$-dependent solution of the field equation
\be\label{2.13}
\frac{\partial\tilde\Gamma_k[\varphi,\psi]}{\partial\varphi^\alpha}
_{|\varphi_k[\psi]}=0\ee
This defines $\varphi_k$ as a $k$-dependent functional of $\psi$.
Since for every scale $k$ the definition (\ref{2.6}) implies
\be\label{2.14}
\frac{\partial\tilde\Gamma_k[\varphi,\psi]}{\partial\varphi^\alpha}=
J_\alpha^*,
\quad \frac{\partial\tilde\Gamma_k[\varphi,\psi]}{\partial\psi^\beta}=
K_\beta^*\ee
and therefore, for $J=0$,
\be\label{2.15}
\frac{d\tilde\Gamma_k[\psi]}{d\psi^\beta}=\frac{\partial\tilde\Gamma_k
[\varphi_k,\psi]}{\partial\psi^\beta}+\frac{\partial
\varphi_k^\alpha[\psi]}
{\partial
\psi^\beta}\frac{\partial\tilde\Gamma_k[\varphi,\psi]}
{\partial\varphi^\alpha}_
{|\varphi=\varphi_k}=K^*_\beta\ee
it is easy to verify that $\tilde\Gamma_k[\psi]$ is the Legendre
transform
of $W_k[K]$ (\ref{2.10}). One concludes for $k\to0$ that $\Gamma_0[\psi]$
is the generating functional for the 1PI Green functions for $\psi$.

We want to employ the flow equation (\ref{2.9}) for finding the
$k$-dependence
of $\Gamma_k[\psi]$. In addition to the corresponding equation for only
one type of fields we have here additional contributions from the
$k$-dependence
of $\Delta_k^{(\varphi)}S$ in (\ref{2.1}). We prefer to
keep\footnote{A different approach \cite{Ell} uses separate cutoff
scales $k$
and $\tilde k$ for the ``gluons'' $\varphi$ and ``quarks'' $\psi$. Letting
$k\to0$ for fixed $\tilde k$ one completely integrates out the gluons.
This
alternative method is most appropriate for massive fields. Then the mass
term acts as an infrared cutoff and prevents the appearance of strong
nonlocalities for $q^2<k^2$}
this additional infrared cutoff for the ``gluons'' in order to maintain
approximate locality of $\Gamma_k[\psi]$ on length scales large compared
to $k^{-1}$.
The evolution equation for
$\Gamma_k[\psi]$ can now be obtained from
a variable transformation which amounts to a shift of $\varphi$
around $\varphi_k[\psi]$
\be\label{2.16}\hat\varphi^\alpha=\varphi^\alpha-
\varphi^\alpha_k[\psi]\ee
Using the general formalism of appendix A one finds
\bear\label{2.17}
&&\frac{\partial}{\partial
t}\Gamma_k[\psi]=\frac{1}{2}\left(\Gamma_k^{(2)}
[\psi]+R_k^{(\psi)}\right)_{\ \ \ \beta'}^{-1\beta}\left\lbrace
\left(\frac{\partial
R_k^{(\psi)}}{\partial t}\right)^{\beta'}_{\ \beta}+
\frac{\partial\varphi^*_{k\alpha'}}{\partial\psi^*_{\beta'}}
\left(\frac{\partial
R_k^{(\varphi)}}{\partial t}\right)^{\alpha'}_{\ \alpha}
\frac{\partial\varphi_k^\alpha}{\partial\psi^\beta}
\right\rbrace\nonumber\\
&&+\frac{1}{2}\varphi^*_{k\alpha'}[\psi]\left(\frac{\partial
R_k^{(\varphi)}}
{\partial t}\right)^{\alpha'}_{\ \ \alpha}\varphi_k^\alpha[\psi]
+\frac{1}{2}\left(\tilde\Gamma^{(2)}_k[\varphi=\varphi_k,\psi]\right)
^{-1\alpha}_{\ \ \ \alpha'}\left(\frac{\partial
R_k^{(\varphi)}}{\partial
t}
\right)^{\alpha'}_{\ \alpha}\ear
It is easy to verify that this equation reduces in the limit
$R_k^{(\varphi)}=0$ to the equivalent of eq. (\ref{2.9}) for fields
$\psi$ only. The corrections in the first two terms involve the explicit
form of the ``classical solution'' $\varphi_k[\psi]$. If one
is interested in
1PI Green functions for $\psi$ with a given number of external legs
one only
needs a polynomial expansion of $\varphi_k[\psi]$ up to a given order.
For example, the evolution of the term $\sim\psi^4$ in $\Gamma_k[\psi]$
needs the classical solution up to the order $\psi^4$ if the series
$\varphi_k[\psi]$ starts with a term quadratic in $\psi$. Additional
knowledge of the form of $\Gamma_k[\hat\varphi,\psi]$ beyond its value
for $\hat\varphi=0$ is only needed for the last correction term
in the form
of
\be\label{2.18}
\left(\hat\Gamma_k^{(2)}[0,\psi]\right)^{\alpha'}_{\ \ \alpha}=
\frac{\partial^2
\Gamma_k[\varphi,\psi]}{\partial\hat\varphi^*_{\alpha'}
\partial\hat\varphi^\alpha}
{}_{|\hat\varphi=0}
+\left(R_k
^{(\varphi)}\right)^{\alpha'}_{\ \ \alpha}\ee
Only the $\psi$-dependence of the effective $\hat\varphi$-propagator plays
a role for the study of 1PI functions for $\psi$.

We finally note that the definition of $\Gamma_k[\psi]$ proposed in this
paper is not the only possible choice of an effective fermionic action.
An alternative formulation is briefly discussed in appendix B.

\section{Integrating out the gauge fields}
\setcounter{equation}{0}
The generalizaton of the discussion of the last section for quarks and
gluons follows ref. \cite{Reu}. All formulae of the last section
remain valid for arbitrary bosonic fields if the indices
$\alpha,\beta$ (\ref{2.2}) include internal indices and Lorentz
indices in addition
to momentum labels\footnote{For a complex field $\varphi$ the sums
are also over negative internal indices $i$ with $\varphi^{-i}
(q)=\varphi^*_i(q)$.}.
If $\psi$ is a Grassmann variable as appropriate for fermions the matrix
$R_k$ in (\ref{2.9}) becomes (cf. appendix C)
\be\label{3.1}
R_k=R_k^{(\varphi)}-R_k^{(\psi)}\ee
Also $\psi^*$ should be replaced by $\bar\psi$ and the index summation
over
$\beta$ should involve both $\psi$ and $\bar\psi$ separately.\footnote{One
may again use negative internal indices for labeling $\bar\psi$, i.e.
$\psi^{-i}
(q)=\bar\psi_i(q)$ and use the summations of the preceding
section including the negative indices.} For the gauge fields we will
choose here a formulation with explicit ghost variables in close
analogy, but slightly different from the formulation in ref. \cite{Reu}.
This
makes our formulation as close as possible to the language of standard
perturbation theory.

We start with the action including a gauge-fixing term in the background
gauge and a corresponding action for the anticommuting ghost fields
$\xi,\bar\xi$
\be\label{3.2}
\hat S[\psi',\xi',a;\bar A]=S[\psi',A']+S_{gf}[a;\bar A]+S_{gh}
[\xi',a;\bar A]\ee
Here $S$ is a gauge invariant functional of the fermion fields $\psi,
\bar\psi$ and the gauge field
\be\label{3.3}
A'_\mu=\bar A_\mu+a_\mu.
\ee
The background gauge field $\bar A_\mu$ appears in the gauge fixing and
ghost terms
\bear
S_{gf}&=&\frac{1}{2\alpha}\int d^d x G^*_z G^z\label{3.4}\\
G^z&=&(D^\mu[\bar A])^z_{\ y}  a^y_\mu\label{3.5}\\
S_{gh}&=&\int d^d x\bar\xi'_y(-D^\mu[\bar A] D_\mu[\bar A+a])^y_{\ z}
\xi^{\prime z}\label{3.6}
\ear
Here $D_\mu[\bar A]$ is the covariant derivative in the adjoint
representation
in presence of  the background gauge field $\bar A$. The generating
functional
for the connected Green functions is defined as usual
\be\label{3.7}
W[\eta,\zeta,K;\bar A]=\int {\cal D}\psi' {\cal D}\bar\psi'{\cal D}\xi'
{\cal D}\bar\xi' {\cal D}a\exp-\{\hat S-\int d^d x[\bar\eta \psi'+
\eta\bar\psi'+\bar\zeta\xi'+\zeta\bar\xi'+Ka]\}\ee
We have introduced here also sources $\zeta$ for the ghost fields and
note that the source $K^\mu_z$ couples to the gauge field fluctuation
$a^z_\mu$ and therefore transforms homogeneously under gauge
transformations
as an adjoint tensor. The $k$-dependent version $W_k$ obtains from $W$ by
adding to $\hat S$ the infrared cutoff piece
\be\label{3.8}
\Delta_k S=\Delta^{(\psi)}_k S+\Delta ^{(A)}_kS+\Delta^{(gh)}_k S
\ee
Here the fermionic cutoff reads
\bear\label{3.9}
\Delta^{(\psi)}_k S&=&\left(\frac{1}{2}\right)\bar\psi'_{\beta'}
(R^{(\psi)}_k)^{\beta'}
_{\ \beta}\psi^{\prime\beta}\nonumber\\
&=&\left(\frac{1}{2}\right)\int d^d x\bar\psi' Z_{\psi,k}
(i\gamma^\mu D_\mu[\bar A])\ r_k^{(\psi)}(-D^2[\bar A]/k^2)\psi'
\ear
with $D_\mu$ the covariant derivative in the appropriate representation
$(D^2=D_\mu D^\mu)$ and $r^{(\psi)}_k$ a dimensionless function. The
factor
$1/2$ is appropriate for Majorana fermions \cite{CWF}, \cite{Ell}. For the
gauge field cutoff we choose
\bear\label{3.10}
\Delta^{(A)}_kS&=&\frac{1}{2} a^*_{\alpha'}(R_k)^{\alpha'}_{\ \alpha}
a^\alpha\nonumber\\
&=& \frac{1}{2} \int d^d xa^y_\nu \left[{\cal D}[\bar A]\ r_k^{(A)}
\left(\frac{{\cal Z}^{-1}_{A,k}{\cal D}[\bar A]}{k^2}\right)\right]
^{y\mu}_{\nu z} a^z_\mu
\ear
with ${\cal D}[\bar A]$ an appropriate operator generalizing a covariant
Laplacian in the adjoint representation which will be explained below.
The matrix ${\cal Z}_{A,k}$ accounts for an appropriate wave function
renormalization.  Finally, we take for the ghosts
\bear\label{3.11}
\Delta^{(gh)}_kS&=&\bar\xi'_{\gamma'}(R^{(gh)}_k)^{\gamma'}_{\ \ \gamma}
\xi^{\prime
\gamma}\nonumber\\
&=&\int d^d x\bar\xi'_y [Z_{gh,k} {\cal D}_s[\bar A] r_k^{(gh)}
({\cal D}_s[\bar A]/k^2)]^y_{\  z}\xi^{\prime z}
\ear
with ${\cal D}_s[\bar A]=- D^2[\bar A]$ in the adjoint representation.
A good choice for the dimensionless function $r_k$ is\footnote{The
function
$r_k^{(\psi)}$ may also be chosen differently from (\ref{3.12}) in order
to
avoid that $R_k$ diverges for vanishing covariant momenta.}
\be\label{3.12}
r_k(y)=\frac{e^{-y}}{1-e^{-y}}\ee
such that
\be\label{3.13}
\lim_{{\cal D}\to 0} R_k=Z_k k^2
\ee
The $k$-dependent functions $Z_{\psi,k,}{\cal Z}_{A,k}$ and $Z_{gh,k}$
will be adapted to corresponding wave function renormalization constants
in the kinetic terms for the fermions, gauge fields and ghosts.
 In principle, they can depend on the background field $\bar A$.
The infrared cutoff piece $\Delta_k S$  cuts off all quantum fluctuations
with covariant momenta smaller than $k$ in the functional integral
defining $W_k$. For covariant momenta larger than $k$ the infrared cutoff
is ineffective and its contribution to the propagator is exponentially
suppressed.

Performing a Legendre transform  and subtracting the IR cutoff piece
again (c.f. (\ref{2.6}), (\ref{2.8})) we arrive at the effective average
action $\Gamma_k[\psi,\xi,A,\bar A]$, where $A=\bar A+\bar a$ and
$\bar a$ is
conjugate to $K$. The dependence of $\Gamma_k$ on the scale $k$
is described by an exact evolution equation analogous to eq. (\ref{2.9}),
with a negative sign for the contributions $\sim R_k^{(\psi)}$ and
$R_k^{(gh)}$. We note that  $\Gamma_k$ only involves terms with an even
number of ghost fields due to the symmetry $\bar\xi'\to -\bar\xi',
\xi'\to-\xi'$ of the  $S_{gh}$ and $\Delta_k^{(gh)} S$. In consequence,
the ghost field equations
\be\label{3.14}
\frac{\delta\Gamma_k}{\delta\bar\xi}=0,\ \frac{\delta \Gamma_k}
{\delta\xi}=0
\ee
have always the solution $\bar\xi=\xi=0$. We therefore can extract the
propagators and vertices for the physical particles from the effective
action for $\bar\xi=\xi=0$:
\be\label{3.15}
\Gamma_k[\psi,A,\bar A]=\Gamma_k[\psi,0,A,\bar A]\ee
Nevertheless, the evolution equation for $\Gamma_k[\psi, A, \bar A]$
obtains
a contribution from the variation of the infrared cutoff of the ghost
fields as given by
\bear
\frac{\partial}{\partial t}\Gamma_k[\psi,A,\bar A]&=&\frac{1}{2}\Tr
\left\{
\left(\frac{\partial}{\partial t} R_k^{(A)}\right)\left(\Gamma_k^{(2)}
+R_k\right)^{-1}\right\}\nonumber\\
&&-\left(\frac{1}{2}\right)\Tr \left\{\left(\frac{\partial}{\partial t}
R_k^{(\psi)}\right)\left(\Gamma_k^{(2)}+R_k\right)^{-1}\right\}-
\varepsilon_k
\label{3.16}\\
\varepsilon_k&=&\Tr\left\{\left(\frac{\partial}{\partial
t}R_k^{(gh)}\right)\left(\Gamma_k^{(gh)(2)}+R_k^{(gh)}\right)^{-1}\right\}
\label{3.17}
\ear
Here $\Gamma^{(2)}_k+R_k$ in (\ref{3.16}) is the matrix of second
functional
derivatives of $\Gamma_k+\Delta_k^{(\psi)} S+\Delta_k^{(A)}S$ with respect
to $\psi$ and $A$ at fixed $\bar A$. For this we have exploited that the
larger matrix of second functional derivatives of  $(\Gamma_k+\Delta_kS)
[\psi,\xi, A, \bar A]$ is block diagonal in the $(\psi, A)$ and $\xi$
components for $\xi=0$.  The ghost dependence of $\Gamma_k[\psi,\xi,
A,\bar A]$
appears in the evolution equation for $\Gamma_k [\psi, A,\bar A]$ only
through
the term $\varepsilon_k$ which involves the second functional derivative
with
respect to the ghost fields $\Gamma_k^{(gh)(2)}$, which is evaluated at
$\bar\xi=\xi=0$ and may depend on $\psi,A,\bar A$.\footnote{In the
formulation of ref. \cite{Reu} the quantity $\Gamma_k^{(gh)(2)}$ is
replaced by $D_s[\bar A]$. For the computation of the evolution of
$\Gamma_k[\psi,A,A]$ performed in ref. \cite{Reu} this is equivalent to
(\ref{3.18}).} We will not  pay much  attention to the detailed form of
$\Gamma_k^{(gh)(2)}$ in the present paper and approximate  it by its
``classical'' value (cf. (\ref{3.6}))
\be\label{3.18}
\Gamma_k^{(gh)(2)}=-D^\mu[\bar A] D_\mu[A].
\ee
In order to complete the formal setup  of our investigation we need to
specify
the operator ${\cal D}$ in eq. (\ref{3.10}).  A good choice is
\be\label{3.19}
{\cal D}[\bar A]=\Gamma_k^{(A)(2)}[\bar A]
\ee
where $\Gamma_k^{(A)(2)}$ is the second functional derivative of
$\Gamma_k[\psi,A,\bar A]$ with respect to $A$ for fixed $\bar A$ and
$\psi=0$, evaluated at the point $A=\bar A$. As in previous formulations
\cite{Reu}, the effective average action $\Gamma_k[\psi,A,\bar A]$ is
gauge
invariant with respect to  simultaneous  gauge transformations of $\psi,
A$ and $\bar A$.

We can now  apply the formalism of the last section in order to
``integrate out'' the gluon fields $A$.
Derivatives with respect to $\varphi$ in the preceding section are
replaced by derivatives with respect to $A$ at fixed background field
$\bar A$.
In particular, the classical field equation, whose solution is $A_k$,
reads (c.f. (\ref{2.13}))
\be\label{3.20}
\frac{\delta\tilde\Gamma_k[\psi,A,\bar A]}{\delta A^z_\mu(x)}_{|A=A_k}=0
\ee
At this point $A_k$ becomes a functional of $\psi$ and $\bar A$. Due to
the covariance of the field equation (\ref{3.20}) we may gauge
transform\footnote{The functional dependence of $A_k$ on $\psi$ and
$\bar A$ is such that a gauge transformation on $\psi$ and $\bar A$
results in a corresponding inhomogeneous gauge transformation of $A_k$.}
any given solution $A_k[\psi,\bar A]$ into a corresponding solution
$A_k[\psi',\bar A']=A'_k$.
In consequence, the effective action
\be\label{3.21}
\Gamma_k^{eff}[\psi,\bar A]=\Gamma_k[\psi,A=A_k[\psi,\bar A],\bar A]
+\Delta^{(A)}_kS[A=A_k[\psi,\bar A],\bar A]
\ee
is gauge-invariant. For our purposes we want to work with an effective
action involving only the quark fields. This requires to fix the
background
field $\bar A$ conveniently. A possible choice is
\bear\label{3.22}
&&\bar A =0\ ,\ A_k[\psi]=A_k[\psi,\bar A=0]\\
&&\Gamma_k[\psi]=\Gamma^{eff}_k[\psi,\bar A=0]=\Gamma_k[\psi,A=
A_k[\psi],\bar A=0]+\Delta_k^{(A)}S[A=A_k[\psi],\bar A=0].\nonumber
\ear
In this version $R_k^{(A)}$ becomes  a simple function of momenta.
Summarizing our adaptation of eq. (\ref{2.17}) for quarks and gluons
and using the
results of appendix C
one obtains\footnote{For general $\bar A$ there is also a purely $\bar A$
dependent contribution to $\Gamma_k[\psi,A,\bar A]$ which arises,
for example,
from the effective infrared cutoff for the ghost fields \cite{Reu}.
This plays no role for the choice $\bar A=0$.}
\bear\label{3.23}
\frac{\partial}{\partial t}\Gamma_k[\psi]&=&-\left(\frac{1}{2}\right)
\left(\Gamma^{(2)}_k[\psi]+R_k^{(\psi)}\right)^{-1\beta}
_{\ \ \  \beta'}\left\{\left(\frac{\partial R_k^{(\psi)}}
{\partial t}\right)^{\beta'}_{\ \beta}
-\frac{\partial A^*_{k\alpha'}}{\partial\bar\psi_{\beta'}}
\left(\frac{\partial R_k^{(A)}}{\partial t}\right)^{\alpha'}_{\ \alpha}
\frac{\partial A^\alpha_k}{\partial  \psi^\beta}\right\}\nonumber\\
&&+\frac{1}{2}\left(\Gamma^{(2)}_k[\psi,A=A_k[\psi],\bar A=0]+R_k^{(A)}
\right)^{-1\alpha}_{\phantom{-1}\  \alpha'}\left(\frac{\partial R_k^{(A)}}
{\partial t}\right)^{\alpha'}_{\ \alpha}\nonumber\\
&&+\frac{1}{2}A^*_{k\alpha'}\left(\frac{\partial R_k^{(A)}}
{\partial t}\right)^{\alpha'}_{\ \alpha} A^\alpha_k -
\varepsilon_k[\psi, A_k[\psi],\bar A=0]
\ear
The factor $(\frac{1}{2})$ in the first term is absent for
Dirac spinors. The flow equation (\ref{3.23})
is the central equation of this paper.

In order to exploit this equation we need $A_k[\psi],\
\Gamma^{(2)}_k[\psi,A_k,0]$ and $\varepsilon_k[\psi,A_k,0]$.
Following ref. \cite{Reu} it is illustrating to divide
$\Gamma_k[\psi,A,\bar A]$ into a gauge invariant effective
action depending only on one gauge field
\be\label{3.24}
\Gamma_k[\psi, A]=\Gamma_k[\psi,A,\bar A=A]
\ee
and a generalized gauge fixing term which vanishes for $\bar A=A$
\be\label{3.25}
\Gamma_k^{gauge}[\psi,A,\bar A]=\Gamma_k[\psi,A,\bar A]
-\Gamma_k[\psi,A].\ee
Besides the classical gauge-fixing term and quantum corrections to
it the part $\Gamma_k^{\rm gauge}$ contains also the $k$-dependent
counterterms, as for example a $k$-dependent gluon mass. (In perturbation
theory this vanishes for $k\to0$.) These counterterms are related
to the gauge-invariant kernel $\Gamma_k[\psi,A]$ by generalized
Slavnov-Taylor identities \cite{Mar}, \cite{Ell2}. We will not
discuss the role of the counterterms very explicitly in the present
paper. (It is conceivable that at some later stage it may become
advantageous to subtract them for an improved definition of $\Gamma_k
[\psi]$.)
Using the explicit form of the gauge invariant infrared
cutoff \cite{Reu}
\be\label{3.26}
\Delta_k^{(A)}S=\frac{1}{2}\int d^dx(A_y^\nu-\bar A_y^\nu)(R_k)^{y\mu}
_{\nu z}(A^z_\mu-\bar A^z_\mu)\ee
the relevant classical field equation (\ref{3.20}) for $A$ becomes
\be\label{3.27}
\frac{\delta\Gamma_k[\psi,A]}{\delta A^z_\mu} +(R_k^{(A)})^{y\mu}_{\nu z}
(A^\nu_y-\bar A^\nu_y)+\frac{\delta\Gamma^{gauge}_k[\psi,A,\bar A]}
{\delta A^z_\mu}=0.\ee
As a consequence of the gauge invariance of $\Gamma_k[\psi,A]$ the gauge
degrees of freedom contained in $A$ only get fixed by the last two terms
in eq. (\ref{3.27}). Similarly, the inverse propagator for the
gauge degrees
of freedom obtains contributions only from $\Gamma^{gauge}_k$
\be\label{3.28}
\Gamma^{(2)}_k[\psi,A,\bar A]=\Gamma^{(2)}_k[\psi,A]+\Gamma^{gauge(2)}_k
[\psi,A,\bar A]\ee
since the contribution from $\Gamma^{(2)}_k[\psi,A]$  vanishes for them.

\section{Effective quark interactions}
\setcounter{equation}{0}

The flow equation (\ref{3.23}) is an exact nonperturbative   equation.
Its solution for $k\to 0$ contains the full information on all 1PI Green
functions for the fields $\psi$. Its solution, however, will only be
approximative and involves a truncation of the general form of
$\Gamma_k[\psi,A,\bar A]$. For a first illustration of the effective
action for quarks $\Gamma_k[\psi]$ we make here the simple ansatz
(with $\psi$ representing Dirac spinors)
\bear\label{4.1}
&&\Gamma_k[\psi,A,\bar A=0]=\int d^4 x\left\{ i Z_{\psi,k}\bar\psi
\gamma^\mu\tilde D_\mu\psi\right.\nonumber\\
&&\left.+\frac{1}{4}\tilde Z_{F,k} F^{\mu\nu}_z F^z_{\mu\nu}+\frac{1}{2}
\tilde Z_{F,k}(\partial_\mu A^\mu)^2+{\cal L}_k[\psi]\right\}
\ear
where ${\cal L}_k[\psi]$ is independent of $A$. Here $F^z_{\mu\nu}$ is the
field strength of a nonabelian $SU(N_c)$ gauge theory and quarks are  in
the
fundamental representation, with
\bear
\tilde D_\mu\psi&=&\partial_\mu\psi-ig_k\tilde Z^{1/2}_{F,k}A^z_\mu
T_z\psi\label{4.2}\\
F^z_{\mu\nu}&=&\partial_\mu A^z_\nu-\partial_\nu A^z_\mu+g_k\tilde
Z^{1/2}_{F,k}
f_{wy}^{\phantom{wy}z} A^w_\mu A^y_\nu\label{4.3}
\ear
The $k$ dependence of $\Gamma_k$ is encoded in the $k$-dependence of
the renormalized
nonabelian gauge coupling $g_k$, the wave function renormalizations
$Z_{\psi,k},\tilde Z_{F,k}$
and ${\cal L}_k[\psi]$. We will discard the $k$-dependence of the
ghost-action and take $\Gamma_{(gh)k}=S_{gh}$ (\ref{3.6}). With this
ansatz an obvious choice for the wave function renormalization in the
effective infrared cutoff is ${\cal Z}_{A,k}=\tilde Z_{F,k}$ and
$Z_{gh,k}=1$. We note that the
truncation (\ref{4.1}) corresponds to a (renormalized) gauge fixing
parameter $\alpha_R=1$. This is convenient for the illustrative purpose
of this section since the classical solution is particularly simple.
For practical computations we remain more
general and suggest the choice $\alpha_R=0$.

With the truncation (\ref{4.1}) the field equation (\ref{3.10}) reads
\be\label{4.4}
\tilde Z_F F^{\mu\nu}_{z\ ;\nu}+ g \tilde Z_F^{1/2} Z_\psi \bar\psi
\gamma^\mu T_z\psi-\tilde Z_F\partial^2 r_k(-\partial^2) A_z^\mu-
\tilde Z_F\partial^\mu\partial_\nu A^\nu_z=0\ee
Here $r_k$ is a dimensionless function of $(-\partial^2/k^2)$
reflecting the details of the infrared cutoff $({\cal D}\equiv
-\partial^2\tilde Z_F)$ and we choose in momentum space (cf. (\ref{3.12}))
\be\label{4.5}
r_k=\frac{\exp\left(-\frac{q^2}{k^2}\right)}{1-\exp\left(-\frac{q^2}
{k^2}\right)}.\ee
Translating to momentum space one obtains (for details see appendix D),
with $\tilde g=g_k\tilde Z^{1/2}_{F,k}$, the field equation
\bear\label{4.6}
&&-\tilde g\frac{Z_\psi}{\tilde Z_F}\int\frac{d^4 p}{(2\pi)^4}\bar\psi
(p)\gamma^\mu T_z\psi(p+q)=\\
&=&P(q) A^\mu_z(q)+i\tilde g f_{zy}^{\phantom{zy}w}\int\frac{d^4 p}
{(2\pi)^4}
A^y_\nu(q-p)\left\{(q^\nu+p^\nu)A^\mu_w(p)\right.
\left.-p^\mu A^\nu_w(p)\right\}\nonumber\\
&&+\frac{\tilde g^2}{2} ( f_{zx}^{\phantom{zx}u}
f_{ywu}+f_{zw}^{\phantom{zw}u} f_{yxu})\int\frac{d^4 p}{(2\pi)^4}
\frac{d^4 p'}{(2\pi)^4} A^{y\mu}(q+p-p') A^x_\nu(-p) A^{w\nu}(p')
\nonumber\ear
where
\be\label{4.7}
P(q)=q^2+q^2r_k=\frac{q^2}{1-\exp(-\frac{q^2}{k^2})}.\ee
If we are interested in a polynomial expansion of $\Gamma_k[\psi]$
we can solve (\ref{4.6}) iteratively. In lowest order one finds
\be\label{4.8}
(A^{(0)}_k(q))^\mu_z=-\tilde g\frac{Z_\psi}{\tilde Z_F}\int \frac{d^4p}
{(2\pi)^4}\bar\psi(p) P^{-1}(q)\gamma^\mu T_z\psi(p+q)
\ee
The next term will involve a correction $\sim\psi^4$ and so forth.
(The term $\sim\psi^4$ can be found in appendix D (D.15)). Inserting
the lowest order classical solution (\ref{4.8}) into the action
(\ref{4.1})
and including the infrared cutoff  $\Delta^{(A)}_k S$ the effective action
for the quarks reads
\be\label{4.9}
\Gamma_k[\psi]=\int \frac{d^4q}{(2\pi)^4}\left\{Z_\psi\bar\psi(q)
\gamma^\mu q_\mu\psi(q)+\tilde{\cal
L}_k[\psi]\right\}+\Gamma_k^{(A)}[\psi]
\ee
with $\tilde{\cal L}$ the Fourier transform of ${\cal L}_k$ and
\be\label{4.10}
\Gamma_k^{(A)}[\psi]=-\frac{1}{2} Z^2_{\psi,k} g^2_k\int
\frac{d^4p_1}{(2\pi)^4}...
\frac{d^4p_4}{(2\pi)^4}(2\pi)^4\delta(p_1+p_2-p_3-p_4)
\frac{1}{P(p_1-p_3)}{\M}\ee
\be\label{4.11}
{\cal M}(p_1,p_2,p_3,p_4)=\left\{\bar\psi^i_a(-p_1)\gamma^\mu(T^z)_i^{\
j}\psi_j^a(-p_3)\right\}\left\{\bar\psi^k_b(p_4)\gamma_\mu(T_z)_k
^{\ \ell}\psi_\ell^b(p_2)\right\}\ee
The curled brackets indicate  contractions over  not explicitly written
indices (here spinor indices), $i,j,k,\ell=1...N_c$ are the colour
indices and $a,b=1...N_f$ the flavour indices of the quarks.
By an appropriate Fierz transformation and using the identity
\be\label{4.12}
(T^z)_i^{\ j}(T_z)_k^{\ \ell}=\frac{1}{2}\delta^\ell_i\delta^j_k-
\frac{1}{2N_c}\delta^j_i\delta^\ell_k
\ee
we can split ${\cal M}$ into three terms \cite{Ell}
\bear
{\cal M}&=&{\cal M}_\sigma+{\cal M}_\rho+{\cal M}_p\label{4.13}\\
{\cal
M}_\sigma&=&-\frac{1}{2}\left\{\bar\psi^i_a(-p_1)\psi_i^b(p_2)\right\}
\left\{\bar\psi^j_b(p_4)\psi_j^a(-p_3)\right\}\nonumber\\
&&+\frac{1}{2}\left\{\bar\psi^i_a(-p_1)\gamma^5
\psi_i^b(p_2)\right\}\left\{
\bar\psi^j_b(p_4)\gamma^5\psi_j^a(-p_3)\right\}\label{4.14}\\
{\cal M}_\rho&=&\frac{1}{4}\left\{\bar\psi^i_a(-p_1)
\gamma_\mu\psi_i^b(p_2
\right\}\left\{\bar\psi^j_b(p_4)\gamma^\mu\psi_j^a
(-p_3)\right\}\nonumber\\
&&+\frac{1}{4}\left\{\bar\psi^i_a(-p_1)\gamma_\mu\gamma^5
\psi_i^b(p_2)\right\}
\left\{\bar\psi^j_b(p_4)\gamma^\mu\gamma^5\psi_j^a(-p_3)\right\}
\label{4.15}\\
{\cal M}_p&=&-\frac{1}{2N_c}\left\{\bar\psi^i_a(-p_1)
\gamma_\mu\psi_i^a(-p_3)
\right\}\left\{\bar\psi^j_b(p_4)\gamma^\mu\psi_j^b(p_2)\right\}
\label{4.16}\ear
In terms of the Lorentz invariants
\bear\label{4.17}
s&=&(p_1+p_2)^2=(p_3+p_4)^2\nonumber\\
t&=&(p_1-p_3)^2=(p_2-p_4)^2\ear
we recognize that the quantum numbers of the fermion bilinears in ${\cal
M}_
\sigma$ correspond to colour singlet, flavour non-singlet scalars in the
$s$-channel and similarly for spin-one mesons for ${\cal M}_\rho$. In
analogy to ref. \cite{Ell} we associate these terms with the scalar mesons
of
the linear $\sigma$-model and with the $\rho$-mesons. The bilinears in the
last term ${\cal M}_p$ correspond to a colour and flavour singlet spin-one
boson in the $t$-channel. These are the quantum numbers of the pomeron.

The effective action $\Gamma_{k_0}[\psi]$ (4.9)
evaluated at some conveniently chosen scale $k_0$ constitutes the
``initial value'' for the subsequent evolution of $\Gamma_k[\psi]$ for
$k<k_0$.
The evolution of $\Gamma_k[\psi]$ with the scale $k$ is then described
by the exact evolution equation (\ref{3.23}). The latter has to
be evaluated
with the approximations (\ref{4.1}). We are interested in the evolution
of the two- and four-point functions for the quarks. The respective
flow equations for these quantities obtain by taking the second and
fourth functional derivative of eq. (\ref{3.23}) at $\psi=\bar\psi=0$.
We label the
different contributions on the r.h.s. of eq. (\ref{3.6}) by
\be\label{4.18}
\frac{\partial}{\partial t}\Gamma_k[\psi]=-\gamma_\psi+\gamma_{A\psi}
+\gamma_A+\gamma_c-\epsilon\ee
and discuss them separately.

The first term
\be\label{4.19}
\gamma_\psi=Tr\left\{\left(\Gamma_k^{(2)}[\psi]+R^{(\psi)}_k\right)^{-1}
\frac{\partial}{\partial t}R_k^{(\psi)}\right\}\ee
is the standard contribution of a pure fermionic theory. It is graphically
represented by a fermion loop in fig. 1. Including only a four-fermion
interaction in $\Gamma_k[\psi]$ the contribution of $\gamma_\psi$ to the
evolution of the two- and four-point functions of the quarks is shown in
fig. 2a and fig. 2b, respectively. In appendix E we expand the
$\psi$-dependent
propagator $(\Gamma_k^{(2)}[\psi]+R_k^{(\psi)})^{-1}$ in the
background quark
fields. From there the contributions to the two- and four-point functions
are easily computed. In this section we will not elaborate further on this
term and rather concentrate on the new ``correction terms'' which arise
from the gluon and ghost fluctuations. More details on $\gamma_\psi$ can
be found in appendix E (cf. (E.14) and (E.17)).

The remaining terms $\gamma_{A\psi},\gamma_A$ and $\gamma_c$
involve $R_k^{(A)}$ and reflect the contributions from gluons, whereas
$\epsilon$ gives the ghost contribution.
The term
\be\label{4.20}
\gamma_A=\frac{1}{2}Tr\left\{\left(\Gamma_k^{(2)}+R_k^{(A)}\right)^{-1}
\frac{\partial}{\partial t}R_k^{(A)}\right\}\ee
involves only a trace over gluonic degrees of freedom and
accounts for the contribution of gluon fluctuations around the
$\psi$-dependent
classical solution (fig. 3). The contribution of $\gamma_A$ to the
fermionic two- and four-point functions is given by the dependence
of $\Gamma_k^{(2)}[\psi,A_k[\psi],0]_{\ \alpha'}^\alpha$ on $\psi$.
Relevant contributions to $\gamma_A$ therefore arise from terms in
$\Gamma_k[\psi,A,0]$ which are either quadratic in $A$ and also depend
on $\psi$ or are cubic or higher-order in $A$. In our
truncation the first sort
of terms is absent and no relevant contribution to $\gamma_A$
would be present for an abelian gauge theory. For nonabelian
gauge theories we get contributions
from the three- and four-gluon vertices in $\Gamma_k[\psi,A]$, i.e.
\be\label{4.21}
\left(\Gamma_k^{(2)}[\psi,A,\bar A=0]\right)^{y\mu}_{\nu z}=
\tilde Z_F\left\lbrace\left(\tilde{\cal
D}_T[A]\right)^{y\mu}_{\nu z}+\left(\tilde{\cal D}_L[0]\right)
^{y\mu}_{\nu z}-\left(\tilde{\cal D}_L[A]\right)^{y\mu}_{\nu z}
\right\rbrace\delta(x-x')\ee
with
\be\label{4.22}
\left(\tilde{\cal D}_T[A]\right)^{y\mu}_{\nu z}=-(\tilde D^2[A])
^y_{\ z}\delta^\mu_\nu+2i\tilde
g\left(T_w\right)^y_{\ z}F^{w\mu}_{\ \nu}\ee
\be\label{4.23}
\left(\tilde{\cal D}_L[A]\right)^{y\mu}_{\nu z}=-(\tilde D_\nu[A])
\tilde D^\mu[A])^y_{\ z}\ee
Here $\tilde D_\mu[A]$ represents the covariant derivative in
the adjoint representation with gauge coupling $\tilde g$
and $F^{\ \mu}_\nu$ is the nonabelian field strength
associated to the gauge field $A$. We observe that $\Gamma_k[\psi,A,0]$ is
invariant under global gauge transformations of $\psi$ and $A$.
The expression
for $\gamma_A$ does not explicitly depend on $\psi$ in our
truncation and $\gamma_A[\psi=0,A_k]$ or $\Gamma_k[0,A_k,0]$
cannot contain a term linear
in $A_k$. There is therefore no contribution from $\gamma_A$ to the
flow equation of the fermionic two-point function.
An estimate of the contribution to the
four-quark interaction from
$\gamma_A$ therefore amounts
to a computation of the gluon contribution to the evolution of the term
quadratic in $A$ in $\Gamma_k[\psi=0,A,\bar A=0]$. This is presented in
appendix F, and the relevant graphs are indicated in figs. 4a, 4b.
Similarly the ghost contribution $\epsilon$ (fig. 5) is (with the
approximation (\ref{3.18})) only a functional of $A$. The ghost
contribution
quadratic in $A$ is also evaluated in appendix F, and its contribution
$\sim\psi^4$ is depicted in fig. 6.

We have computed $\gamma_A-\epsilon$ in the truncation (\ref{4.1}).
Neglecting a contribution with a different fermionic index structure (the
term $\sim\frac{\partial}{\partial t}H_A(q)$ in appendix F) and
another contribution $\sim \frac{\partial}{\partial t}
\ln \tilde Z_F$ from the
wave function renormalization $\tilde Z_F$ in $R_k$ one obtains (F.56)
\bear\label{4.24}
&&\gamma_A-\epsilon=\frac{1}{2}N_cg^4_kZ^2_\psi\int\frac{d^4p_1}{(2\pi)^4}
...\frac{d^4p_4}{(2\pi)^4}(2\pi)^4\delta(p_1+p_2-p_3-p_4)P^{-2}(p_1-p_3)
\nonumber\\
&&{\cal G}(p_1-p_3){\cal M}(p_1,p_2,p_3,p_4)\nonumber\\
&&{\cal G}(q)=\frac{k^2}{16\pi^2}\left\lbrace4+12\frac{k^2}
{q^2}-8\frac{k^4}{(q^2)^2}
-\exp\left(-\frac{q^2}{2k^2}\right)\left(9+8\frac{k^2}{q^2}-8\frac{k^4}
{(q^2)^2}\right)\right\rbrace\ear
The gluon and ghost fluctuations induce therefore a running of the
coefficient function multiplying the term $\sim {\cal M}$ in (\ref{4.9}).
More precisely, we may parametrize
\bear\label{4.25}
&&\Gamma_{k,4}[\psi]=-Z_\psi^2\int\frac{d^4p_1}{(2\pi)^4}...
\frac{d^4p_4}{(2\pi)^4}(2\pi)^4\delta(p_1+p_2-p_3-p_4)\nonumber\\
&&\left\lbrace\lambda_m(p_1,p_2,p_3,p_4){\cal M}(p_1,p_2,p_3,p_4)+\quad
{\rm other\ index\ structures}\right\rbrace\ear
Then $\gamma_A-\epsilon$ induces for $k<k_o$ a deviation of $\lambda_m$
from its ``classical value''
\be\label{4.26}
\lambda_m^{(c)}(p_1,p_2,p_3,p_4)=\frac{1}{4}g^2_k\left
(P^{-1}(p_1-p_3)+P^{-1}
(p_2-p_4)\right)\ee
as given by the contribution to the flow equation
\be\label{4.27}
\frac{\partial}{\partial t}\lambda_m[\gamma_A-\epsilon]
=-\frac{1}{4}N_cg_k^4\left\lbrace P^{-2}(p_1-p_3){\cal G}(p_1-p_3)+P^{-2}
(p_2-p_4){\cal G}(p_2-p_4)\right\rbrace\ee
Therefore the gluon contribution to $\frac{\partial}{\partial t}\lambda_m$
depends only on $t$ and vanishes for $t\gg k^2$. Only the behaviour
of the four-point function for $t$ of the order of $k^2$ or smaller
gets modified. Details of the behaviour of ${\cal G}(q)$ for $q^2\ll k^2$
can
be found in appendix F.

We also observe that
$\gamma_A$ and $\epsilon$
account for the gluon and ghost contributions to  the effective
gluon propagator, whereas  the contribution from quark loops is implicitly
contained in $\gamma_\psi$ (fig. 2b). We note that the latter is not
distinguished any more from any other fermionic contributions, as, for
example, from an explicit four-quark interaction in ${\cal L}_k[\psi]$.
Only
in lowest order in standard perturbation theory the contribution from
fig. 2b corresponds exactly to the quark contribution to
the renormalized gluon
propagator.

The contribution $\gamma_{c}$ is quadratic in  the classical
solution $A_k[\psi]$ and gives a contribution
to the four-point function
\be\label{4.28}
\gamma_{c}=\frac{1}{2}g_k^2Z^2_\psi\int\frac{d^4p_1}{(2\pi)^4}...
\frac{d^4p_4}{(2\pi)^4}(2\pi)^4\delta(p_1+p_2-p_3-p_4)K(p_1-p_3){\cal M}
(p_1,p_2,p_3,p_4)\ee
with
\be\label{4.29}
K(q)=q^2P^{-2}(q)\left(\frac{\partial}{\partial t}r_k(q)-\tilde\eta_F
r_k(q)\right)\ee
and
\be\label{4.30}
\tilde\eta_F=-\frac{\partial}{\partial t}\ln \tilde Z_F\ee
This again results in a contribution to the running of $\lambda_m$
\be\label{4.31}
\frac{\partial}{\partial t}\lambda_m[\gamma_c]=-\frac{1}{4}g^2_k
\left(K(p_1-p_3)+K(p_2-p_4)\right)\ee
Comparison with eq. (4.26) shows that this contribution accounts
exactly for the $k$-dependence of the infrared cutoff $R_k$
contained in $P$. Taking only this contribution into acount the solution
of the flow equation therefore describes the $k$-dependence of $P^{-1}$,
whereas $g^2_{k_0}$ is not modified at this level. The effective running
of $g^2_k$ arises through the contributions
$-\gamma_\psi+\gamma_A-\epsilon
+\gamma_{A\psi}$.

The last contribution $\gamma_{A\psi}$ finally involves the explicit
$\psi$
dependence of the classical solution in form of the term $\sim \partial
A_k/\partial \psi$. It is graphically represented in fig. 7
and reads
\bear\label{4.32}
&&\gamma_{A\psi}=-\int\frac{d^4q}{(2\pi)^4}\tilde Z_Fq^2
\left(\frac{\partial}{\partial t}r_k(q)-\tilde\eta_Fr_k(q)\right)
\nonumber\\
&&\int\frac{d^4p}{(2\pi)^4}\frac{d^4p'}{(2\pi)^4}\frac{\delta
A^z_{k\mu}(q)}
{\delta\psi^a_i(p)}\left(\frac{1}{\Gamma_k^{(2)}[\psi]+
R_k^{(\psi)}}\right)^{aj}_{bi}(p,p')\frac{\delta
A^\mu_{kz}(-q)}{\delta\bar\psi^j_b(p')}\ear
In lowest order we insert (\ref{4.8}) for the classical solution
\bear\label{4.33}
&&\frac{\delta A^z_{k\mu}(q)}{\delta\psi^a_i(p)}=\tilde g
\frac{Z_\psi}{\tilde Z_F}P^{-1}(q)\bar\psi^l_a(p-q)\gamma_\mu(T^z)_l^{\
i}\nonumber\\
&&\frac{\delta A^\mu_{kz}(-q)}{\delta\bar\psi^j_b(p')}=
-\tilde g\frac{Z_\psi}{\tilde Z_F}P^{-1}(q)\gamma^\mu(T_z)
_j^{\ l}\psi^b_l(p'-q)\ear
and we expand
\be\label{4.34}
\Gamma_k^{(2)}[\psi]=\Gamma_k^{(2)}[0]+\Gamma_k^{(4)}[0]
\bar\psi\psi+...\ee
\bear\label{4.35}
\left(\Gamma_k^{(2)}[0]\right)^{aj}_{bi}&=&Z_\psi q\llap/\delta^a_b
\delta^j_i(2\pi)^4\delta(q-q')
\nonumber\\
&&+(L_k^{(2)}[0])^a_b(q)\delta^j_i(2\pi)^4\delta(q-q')\ear
(Here $L_k^{(2)}[0]$ is a possible contribution from ${\cal L}_k[\psi]$
in
eq. (\ref{4.1}), as for example a quark mass term.) The precise
index structure
of $\Gamma_k^{(4)}[0]\bar\psi\psi$ can be found in the appendix E (cf.
E.4).
Expanding $(\Gamma_k^{(2)}+R_k)^{-1}$ and assuming that $L_k^{(2)}[0]$
is flavour-diagonal
\be\label{4.36}
\left(L_k^{(2)}[0]\right)^a_b(q)=Z_\psi m_a(q)\delta^a_b\gamma^5\ee
one obtains the following contribution to the evolution of the
two-point function (fig. 7):
\bear\label{4.37}
&&\gamma^{(2)}_{A\psi}=\frac{N_c^2-1}
{N_c}Z_\psi g^2_k\int\frac{d^4q}{(2\pi)^4}
\int\frac{d^4p}{(2\pi)^4}(q+p)^2\left(\frac{\partial}{\partial t}r_k(q+p)
-\tilde\eta_Fr_k(q+p)\right)
\nonumber\\
&&P^{-2}(q+p)\sum_a\bar\psi_a(p)\frac{q\llap/(1+r_k^{(\psi)}(q))
-2m_a(q)\gamma^5}
{q^2(1+r_k^{(\psi)}(q))^2+m^2_a(q)}\psi^a(p)\ear
Two similar contributions to the four-point function arise if we
include the terms $\sim\bar\psi\psi\psi$ in $\partial
A_k/\partial\bar\psi$
and
similar in $\partial A_k/\partial \bar\psi$ (fig. 8a). Another
contribution
to the four-point function arises through the expansion of $\Gamma_k^{(2)}
[\psi]$ (4.34) as shown in fig. 8b. We observe that in
perturbation theory the graph 8a corresponds to a contribution to the
renormalization of the $\bar\psi\psi A_k$ vertex (fig. 9a) and similarly
the correspondence of the graph 8b is given in fig. 9b.

To summarize, all the contributions to the flow equation (4.18)
have a simple
well-defined meaning, which can easily be expressed in terms of graphs
with rules given in fig. 10. For small gauge coupling there is a one
to one correspondence with associated pieces in standard perturbation
theory. Our flow equation is, however, not limited to the perturbative
regime.

\section{Heavy quark approximation}
\setcounter{equation}{0}

In the limit of infinitely large quark masses our formalism simplifies
considerably. If the momenta in the $n$-point functions remain
bounded (and therefore much smaller than the heavy quark mass),
we can omit in eq. (\ref{3.23}) the terms involving the
inverse fermion propagator $(\Gamma_k^{(2)}[\psi]+R_k^{(\psi)})^{-1}$.
Their contribution is suppressed by inverse powers of the
quark masses.
In the language of the last section this results to
$\gamma_\psi=0,\gamma_{A\psi}
=0$. The remaining computation amounts to an investigation of the
pure gluon
theory with static quarks. This is done most easily in the language
where the gluon fields are kept explicitly and the relevant effective
action is $\Gamma_k[\psi,A,\bar A=0]$. It is instructive, however, to
understand at every step the exact equivalence with the effective
quark theory developed in the last sections.

If one wants to extract the effective four-quark interaction which
encodes the
effective heavy quark potential, one needs the $k$-dependent
effective gluon
propagator and the effective vertex $\bar\psi\psi A$.
We first describe  here (for arbitrary quark masses) the general
framework how an effective four-quark interaction
obtains from ``gluon
exchange'' in the formulation where both quark and gluon
degrees of freedom
are kept explicitly. We then specialize to the heavy quark limit and
show the equivalence with the formulation in terms of only quark
degrees of freedom. The inverse gluon propagator
\be\label{5.1}
\left(\Gamma_k^{(2)}[\psi=0,A=0,\bar A=0]\right)^{y\mu}_{\nu z}
(q)=(G_A(q)\delta^\mu_\nu+H_A(q)q_\nu q^\mu)\delta^y_z\ee
specifies the term quadratic in $A$
\be\label{5.2}
\Gamma_{k,2}^{(A)}=\frac{1}{2}\int\frac{d^4q}{(2\pi)^4}A^\nu_y(-q)
(\Gamma_k^{(2)}[\psi=0,A=0,\bar A=0])^{y\mu}_{\nu z}A^z_\mu(q)\ee
whereas the quark-gluon vertex is encoded in
\be\label{5.3}
\Gamma_k^{(\bar\psi\psi A)}=\int\frac{d^4p}{(2\pi)^4}\frac{d^4q}
{(2\pi)^4}\bar\psi^i_a(p)G_\psi(p,q)
\gamma^\mu(T_z)^{\ j}_i\psi^a_j(p+q)A^z_\mu(-q)\ee
Knowledge of $G_A, H_A$ and $G_\psi$ permits to compute the classical
solution  $A_k$ in order $\bar\psi\psi$
\be\label{5.4}
(A_k^{(0)}(q))^\nu_z=-S^\nu_\mu(q)\int\frac{d^4p}{(2\pi)^4}
\bar\psi_a^i(p)G_\psi(p,q)\gamma^\mu(T_z)_i^{\ j}\psi^a_j(p+q)\ee
Here $S=(\Gamma_k^{(2)}[0]+R_k)^{-1}$ is the gluon propagator in presence
of the infrared cutoff
\be\label{5.5}
(R_k^{(A)})^{y\mu}_{\nu z}(q)=(R_k(q)\delta^\mu_\nu+
\tilde R_k(q)q_\nu q^\mu)\delta^y_z\ee
and reads
\bear\label{5.6}
&&S^\nu_\mu(q)=(G_A(q)+R_k(q))^{-1}\left\lbrace\delta^
\nu_\mu-q^\nu q_\mu(H_A(q)+\tilde R_k(q))\cdot\right.\nonumber\\
&&\left.[G_A(q)+R_k(q)+q^2(H_A(q)+\tilde R_k(q))]^{-1}\right\rbrace\ear
In general, the reduced three-point function $G_\psi(p,q)$ may
still involve Dirac matrices (e.g. terms $\sim\gamma^\nu p_\nu$).
In the heavy quark limit considered here we can neglect this
possibility and treat $G_\psi$ as a scalar function.
Inserting the classical solution into (\ref{5.2}) and (\ref{5.3}) and
accounting for the term $\Delta_k^{(A)}S$ (\ref{3.22}) yields
the effective quark four point function
\bear\label{5.7}
&&\Gamma_{k,4}^{(\psi)}=-\frac{1}{2}\int\frac{d^4p}{(2\pi)^4}
\frac{d^4p'}{(2\pi)^4}\frac{d^4q}{(2\pi)^4}S^\nu_\mu(q)
G_\psi(p,q)G_\psi(p',-q)
\nonumber\\
&&\lbrace\bar\psi^i_a(p)\gamma^\mu(T_z)_i^{\ j}\psi^a_j(p+q)
\rbrace\lbrace
\bar\psi^k_b(p')\gamma_\nu(T^z)_k^{\ l}\psi^b_l(p'-q)\rbrace
\nonumber\\
&&=-\frac{1}{2}\int\frac{d^4p_1}{(2\pi)^4}...\frac{d^4p_4}
{(2\pi)^4}(2\pi)^4
\delta(p_1+p_2-p_3-p_4)\cdot\nonumber\\
&&\left\lbrace F_1(p_1,p_2,p_3,p_4){\cal M}(p_1,p_2,p_3,p_4)
+F_2(p_1,p_2,p_3,p_4){\cal N}(p_1,p_2,p_3,p_4)\right\rbrace\ear
with
\bear\label{5.8}
F_1&=&G_\psi(-p_1,p_1-p_3)G_\psi(p_4,p_2-p_4)(G_A(p_1-p_3)+
R_k(p_1-p_3))^{-1}
\nonumber\\
F_2&=&G_\psi(-p_1,p_1-p_3)G_\psi(p_4,p_2-p_4)(H_A(p_1-p_3)+
\tilde R_k(p_1-p_3))\nonumber\\
&&(G_A(p_1-p_3)+R_k(p_1-p_3))^{-1}[G_A(p_1-p_3)+R_k(p_1-p_3)\nonumber\\
&&+(p_1-p_3)^2(H_A(p_1-p_3)+\tilde R_k(p_1-p_3))]^{-1}\ear
Here ${\cal M}$ is given by (\ref{4.13}) and
\bear\label{5.9}
{\cal N}(p_1,p_2,p_3,p_4)&=&\lbrace\bar\psi^i_a(-p_1)
(p\llap/_1-p\llap/_3)
(T_z)_i^{\ j}\psi^a_j(-p_3)\rbrace\nonumber\\
&&\lbrace\bar\psi^k_b(p_4)(p\llap/_2-p\llap/_4)
(T^z)_k^{\ l}\psi^b_l(p_2)\rbrace\ear
We observe that in the heavy quark approximation
the coefficients of the quark
interactions in the $\sigma,\rho$ and pomeron channel (\ref{4.13}) are all
given by the same function $F_1$. For the general gluon propagator
discussed
in this section there is also an additional four-quark interaction
$\sim {\cal N}$ which vanishes in the approximation $H_A=\tilde R_k=0$
used in sect. 4.  The general
quark bilinear is conveniently parametrized by the real functions
$Z_\psi(q)$
and $\bar m_a(q)$
\be\label{5.10}
\Gamma_{k,2}^{(\psi)}=\sum_a\int\frac{d^4q}{(2\pi)^4}\bar\psi_a^i(q)
(Z_\psi
(q)\gamma^\mu q_\mu+\bar m_a(q)\gamma^5)\psi^a_i(q)\ee
The $k$-dependence of the functions $G_A,H_A,G_\psi,Z_\psi$ and $\bar m_a$
relevant for the two- and four-point functions for the quarks can now
be studied using the evolution equation (\ref{3.16}) for $\Gamma_k
[\psi,A,\bar
A=0]$.

In the truncation where only the terms (\ref{5.2}),(\ref{5.3})
and (\ref{5.10}) are kept, it is easy to see that the contributons to the
$k$-dependence of $G_\psi,Z_\psi$ and $\bar m_a$ all involve quark
propagators. In the heavy quark limit
with fixed external momenta they can therefore be neglected and
only the $k$-dependence of $\Gamma_{k,2}^{(A)}$ needs to be considered.
For $Z_\psi$ and $G_\psi$ we may take appropriate momentum-independent
``short-distance couplings''
\bear\label{5.11}
Z_\psi(q)&=&1\nonumber\\
G_\psi(p,q)&=&\tilde Z_F^{\frac{1}{2}}(m_\psi)g(m_\psi)\ear
with renormalized gauge coupling $g$ taken at the scale $k=m_\psi$
and $m_\psi$ the heavy  quark mass. We also may identify $k=m_\psi$
with the
``ultraviolet cutoff'' or the scale where the initial values for the flow
equation are specified, i.e.
\bear\label{5.12}
\tilde Z_F(m_\psi)&=&1\nonumber\\
G_A(q;k=m_\psi)&=&q^2\ear
Solving the flow equation for $G_A(q)$ for $k\to0$
one can compute the heavy
quark four-point function for momenta much smaller than the quark mass.
For $\alpha_R=0$ (see appendix F) it is fully determined by
\bear\label{5.13}
&&\Gamma_{0,4}^{(\psi)}=-\frac{1}{2}g^2(m_\psi)\int\frac{d^4p_1}
{(2\pi)^4}...
\frac{d^4p_4}{(2\pi)^4}(2\pi)^4\delta(p_1+p_2-p_3-p_4)\nonumber\\
&&\lim_{k\to0}G^{-1}_A(p_1-p_3)\left\lbrace{\cal M}(p_1,p_2,p_3,p_4)
+\frac{1}{(p_1-p_3)^2}{\cal N}(p_1,p_2,p_3,p_4)\right\rbrace\ear
For example, a potential with a Coulomb term and
a linear term with string tension $\lambda$ would correspond
in (\ref{5.13}) to
\be\label{5.14}
G_A^{-1}(q)=\frac{1}{q^2}+
\frac{16\pi\lambda}{g^2(m_\psi)(q^2)^2}\ee
The flow equation for $G_A$ is computed in appendix F.

Let us next illustrate the equivalence with the language of
sects. 3 and 4. Neglecting terms $\sim (\bar\psi\psi)^3$ or higher powers
of fermion bilinears, the $k$-dependence of $\Gamma_k[\psi]$ is easily
understood in the heavy quark limit: We insert in
\bear\label{5.15}
&&\frac{\partial}{\partial t}\Gamma_k[\psi]=\frac{\partial}{\partial
t}\Gamma_{k,2}^{(\psi)}[\psi]+\frac{\partial}{\partial
t}\Gamma_{k,2}^{(A)}[A]_{|A=A_k}+\frac{\partial}{\partial t}\Gamma_k
^{(\bar\psi\psi A)}[\psi,A]_{|A=A_k}\\
&&+\frac{\delta \Gamma_{k,2}^{(A)}}{\delta A^z_\mu(q)}_{\Bigl|_{A=A_k}
\Bigr.}\frac{\partial}{\partial t}(A_k^{(0)}(q))^z_\mu+\frac{\delta
\Gamma_k^{(\bar\psi\psi A)}}{\delta
A^z_\mu(q)}_{|A=A_k}\frac{\partial}{\partial t}(A_k^{(0)}
(q))^z_\mu+\frac{\partial}{\partial t}\Delta^{(A)}_kS_{|A=A_k}
\nonumber\ear
the identity for the classical solution
\be\label{5.16}
\frac{\delta\Gamma_k}{\delta A(-q)}_{|A=A_k}=-\frac{\delta\Delta^{(A)}_kS}
{\delta A(-q)}_{|A=A_k}=-R_k^{(A)}(q)A_k(q)\ee
and obtain
\be\label{5.17}
\frac{\partial}{\partial t}\Gamma_k[\psi]=\frac{\partial}{\partial
t}\Gamma_{k,2}^{(\psi)}[\psi]+\frac{\partial}{\partial t}
\Gamma_k^{(\bar\psi\psi A)}
[\psi,A_k]
+\frac{\partial}{\partial t}\Gamma^{(A)}_{k,2}[A_k]+\frac{1}{2}A_k
\frac{\partial R_k^{(A)}}{\partial t}A_k\ee
The first two terms on the r.h.s. of (\ref{5.17})
do not contribute in the heavy quark limit
with fixed external momenta, the third term corresponds
exactly to $\gamma_A-\epsilon$ and the last term to $\gamma_c$. These
remarks generalize, of course, to higher $n$-point functions for the
quarks.

As a side remark
it should be noticed that for given vertices and propagators
a computation of $\gamma_A-\epsilon+\gamma_c$ as a functional of
$A_k$ is independent of the masses of the quarks.
It amounts to a computation within the pure gluon theory without
any explicit
reference\footnote{This can be generalized for a fermion field-dependent
gluon propagator if terms $\sim \bar\psi\psi A^2$ are included.} to the
fermions. It is therefore valid for arbitrary quark masses. In
consequence, the results
for $\gamma_A-\epsilon$ reported in appendix F can  be used
directly for the r.h.s. of the evolution equation (\ref{4.18})
for arbitrary
quark masses. Of course, the expression of $A_k$ as a functional of
$\psi$ involves
the fermionic part of the average action and therefore depends on the
quark
gluon couplings. Also, all vertices and propagators on the r.h.s. of the
flow equation have to be taken for the full theory with light fermions.

We conclude that the flow equation (\ref{3.23}) becomes very simple
if the quark mass is large compared to all momenta in the
Green functions. This static limit is, however, not yet the full
answer to physical questions like the decription of heavy quark
scattering or the heavy quark potential. For these purposes the
momenta $p_1...p_4$ in the heavy quark four-point function have to
be continued analytically to Minkowski space and should be taken
on-shell, i.e. $p_1^2=p_2^2=p^2_3=p_4^2=-m^2_\psi$. They
can therefore not be considered as small compared to $m_\psi$.
In particular, one needs in (5.8) the three-point functions
$G_\psi(p,q)$ for $p^2=(p+q)^2=-m_\psi^2$ in addition to $G_A(q)$.
The flow of this vertex function can again be computed in the
formulation with quarks and gluons. The correspondence in the
pure fermionic flow equations is straightforward and results
in a flow equation for the four-point function for momenta on
mass shell. By Lorentz symmetry $G_\psi(p,q)$ can only depend
on the three independent invariants $p^2,(p+q)^2$ and
$q^2$. In consequence, the on-shell vertex is only a function
of $q^2$
\be\label{5.18}
G_\psi(p,q)_{|p^2=(p+q)^2=-m^2_\psi}\equiv G_\psi(q)\ee
For on-shell situations we therefore should replace in (\ref{5.13})
$g^2(m_\psi)G_A^{-1}(q)$ by\\
$G_\psi^2(q)G^{-1}_A(q)=F(q)$ where
$F(q)=\lim_{k\to0}F_k(q)=\lim_{k\to0}F_1(q)_{|_{\rm on\ shell}}$.The
relevant flow equation for $F_k(q)$
\be\label{5.19}
\frac{\partial}{\partial t}F_k^{-1}(q)=
\frac{\frac{\partial}{\partial
t}G_A(q)}{G^2_\psi(q)}-2\frac{\frac{\partial}{\partial
t}G_\psi(q)G_A(q)}{G^3_\psi(q)}\ee
involves also the flow equation for $G_\psi(q)$. We also note that for
on-shell quarks $(\slp+m_\psi\gamma^5)\psi(p)=0,\ \bar\psi
(p)(\slp+m_\psi\gamma^5)=0$ and therefore $(\cal N)$ (\ref{5.9})
vanishes. Heavy quark scattering is entirely described
by $F(q)$. The heavy quark potential can be extracted from this
function by a three-dimensional Fourier transform.

\section{Conclusions}
\setcounter{equation}{0}
In this paper we have developed a new formalism of how to
integrate out gluons in QCD. The result is an effective
action for quarks with an infrared cutoff $k$ - the average
action for quark fields $\Gamma_k[\psi]$. Only gluons with
momenta $q^2\gsim k^2$ are included in the computation of
$\Gamma_k[\psi]$. An exact non-perturbative evolution equation
describes the dependence of $\Gamma_k$ on the infrared cutoff $k$.
Additional gluon fluctuations are included as $k$ is lowered
and lead to ``residual gluon corrections'' to the purely
fermionic flow equation. Our treatment permits a smooth
transition from a description in terms of quarks and gluons
(appropriate for perturbative QCD) to effective models
involving only quarks. For many practical purposes one is
interested in the limit $k\to0$ where all quantum fluctuations
are (formally) included.

An approximate solution of the flow equations needs truncations
of the most general form of $\Gamma_k$. This restricts the
present range of practical applicability of our formalism. With
the present rough truncations in the gluon sector we see two
areas where this method may lead to interesting new results:
One is colour-neutral strong interaction physcis, in particular the
properties of mesons. The other concerns heavy quark physics
at intermediate distances or momentum scales, say $q^2\approx((0.3-1)
\ {\rm GeV})^2$. On the other hand, present truncations seem
inappropriate for a quantitative treatment of confinement. Here
the low momentum properties of the ``glue'' need to be understood
and, most probably, additional composite degrees of freedom
should be introduced.

This paper is devoted to the development of the
new formalism. Sect. 2 is kept rather general and the general
field-theoretical setting can be applied to problems of integrating
out degrees of freedom in a wide context. In the following sections
we focus on QCD. In view of the length of this paper we
have not yet addressed here
practical applications, which will involve numerical
solutions of truncated flow equations. Nevertheless,
the necessary analytical
computations for this purpose have already largely
been carried out and are presented in the appendices: The heavy
quark potential involves the flow equation for the gluon
propagator which is computed in appendix F. The residual gluon
corrections for the flow equations for light quarks need in
addition the results of appendix E. With the help of these
formulae the way for numerical solutions of the truncated flow
equations seems open and we hope that interesting quantitative
results for QCD at intermediate scales and meson physics will
emerge.

\section*{Appendix A: Field transformations in the flow equation}
\renewcommand{\theequation}{A.\arabic{equation}}
\setcounter{equation}{0}

For a general $k$-dependent nonlinear variable transformation $\hat\varphi
=\hat\varphi_k[\varphi,\psi],\ \hat\psi=\hat\psi_k[\varphi,\psi]$ the flow
equation (\ref{2.9}) transforms into
\bear\label{A.1}
&&\frac{\partial}{\partial
t}\Gamma_k[\hat\varphi,\hat\psi]=\frac{\partial}
{\partial t}\Gamma_{k_{|\varphi,\psi}}-\frac{\partial\Gamma_k
[\hat\varphi,\hat\psi]}
{\partial\hat\varphi^\alpha}
\frac{\partial\hat\varphi_k^\alpha}{\partial t}_{|\varphi,\psi}-
\frac{\partial\Gamma_k[\hat
\varphi,\hat\psi]}{\partial\hat\psi^\beta}\frac{\partial\hat\psi^\beta_k}
{\partial t}_{|\varphi,\psi}\nonumber\\
&&=\frac{1}{2} Tr\lbrace(\tilde\Gamma_k^{(2)})^{-1}
\frac{\partial R_k}{\partial
t}\rbrace
-\frac{\partial\tilde\Gamma_k[\hat\varphi,\hat\psi]}{\partial\hat\varphi
^\alpha}\frac{\partial\hat\varphi_k^\alpha}{\partial t}-
\frac{\partial\tilde
\Gamma_k[\hat\varphi,\hat\psi]}{\partial\hat\psi^\beta}
\frac{\partial\hat\psi
^\beta_k}{\partial t}\nonumber\\
&&+\frac{\partial\Delta_k^{(\varphi)}S[\hat\varphi]}
{\partial\hat\varphi^\alpha}
\frac{\partial\hat\varphi_k^\alpha}{\partial t}+\frac{\partial
\Delta_k^{(\psi)}
S[\hat\psi]}{\partial\hat\psi^\beta}\frac{\partial\hat\psi_k^\beta}
{\partial t}
\ear
(Here $\varphi=\varphi_k[\hat\varphi,\hat\psi]$, the partial derivative
$\partial\hat\varphi^\alpha_k/\partial t$ is taken for fixed
$\varphi$ and
$\psi$ and similar for $\partial\hat\psi^\beta_k/\partial t$.)
The corresponding matrix of second functional derivatives
$\tilde\Gamma_k^{(2)}$
reads now, with combined fields
\be\label{A.2}
\sigma^\gamma=(\varphi^\alpha,\psi^\beta),\ \hat\sigma^\gamma=(\hat
\varphi^\alpha,\hat\psi^\beta),\ee
\be\label{A.3}
(\tilde\Gamma_k^{(2)})^\gamma_{\ \delta}
=\frac{\partial^2\tilde\Gamma_k}{\partial
\hat\sigma_{\gamma'}^*\partial\hat\sigma^{\delta'}}
\frac{\partial\hat\sigma^*_
{\gamma'}}{\partial\sigma^*_\gamma}\frac{\partial\hat\sigma^{\delta'}}
{\delta\sigma^\delta}
+\frac{\partial\tilde\Gamma_k}{\partial\hat\sigma^{\delta'}}
\frac{\partial^2
\hat\sigma^{\delta'}}{\partial\sigma^*_\gamma\partial\sigma^\delta}\ee
We can therefore write
\be\label{A.4}
Tr\lbrace(\tilde\Gamma_k^{(2)})^{-1}\frac{\partial R_k}{\partial t}\rbrace
=\hat G^{\gamma'}_{\ \delta'}\frac{\partial\sigma_\delta^*}{\partial\hat
\sigma_{\delta'}^*}\left(\frac{\partial R_k}
{\partial t}\right)^\delta_\gamma
\frac{\partial\sigma^\gamma}{\partial\hat\sigma^{\gamma'}}\ee
with $\hat G$ the inverse of the ``covariant second functional
derivative''
\be\label{A.5}
\left(\hat\Gamma_k^{(2)}\right)^{\gamma'}_{\ \delta'}
=\Gamma_{k;\delta'}^{\ \ \gamma'}=\frac{\partial^2
\tilde\Gamma_k[\hat\sigma]}
{\partial\hat\sigma^*
_{\gamma'}\partial\hat\sigma^{\delta'}}+
\omega^{\gamma'\ \eta'}_{\ \delta'\ }\frac{\partial
\tilde\Gamma_k[\hat\sigma]}
{\partial\hat\sigma^{\eta'}}
\ee
and ``connection''
\be\label{A.6}
\omega^{\gamma'\ \eta'}_{\ \delta'\ }=\frac{\partial\sigma^*_\gamma}
{\partial
\hat\sigma^*_{\gamma'}}\frac{\partial^2\hat\sigma^{\eta'}}
{\partial\sigma^*_\gamma\partial\sigma^\delta}
\frac{\partial\sigma^\delta}
{\partial\hat\sigma^{\delta'}}\ee
Inserting the special transformation (\ref{2.16})
\bear\label{A.7}
&&\psi=\psi_k[\hat\varphi,\hat\psi]=\hat\psi\nonumber\\
&&\varphi=\varphi_k[\hat\varphi,\hat\psi]=\varphi_k[\hat\psi]+
\hat\varphi\ear
one obtains

\bear\label{A.8}
&&\frac{\partial}{\partial t}\Gamma_k[\hat\varphi,
\psi]=\frac{1}{2}\left(\hat\Gamma_k^{(2)}\right)^{-1\beta}_{\ \ \ \beta'}
\left(
\frac{\partial R_k^{(\psi)}}{\partial t}\right)^{\beta'}_{\ \beta}
+\frac{1}{2}\left(\hat\Gamma_k^{(2)}\right)^{-1\alpha}
_{\ \ \  \alpha'}\left(
\frac{\partial R_k^{(\varphi)}}{\partial t}\right)^{\alpha'}_{\ \alpha}
\nonumber\\
&&+\frac{1}{2}\left(\hat\Gamma_k^{(2)}\right)^{-1\alpha}_{\ \ \ \beta}
\frac{\partial\varphi^*_{k\alpha'}}{\partial\psi^*_\beta}
\left(\frac{\partial
R_k^{(\varphi)}}{\partial t}\right)^{\alpha'}_{\ \alpha}+
\frac{1}{2}\left(\hat\Gamma_k^{(2)}\right)^{-1\beta}_{\ \ \ \alpha'}
\left(\frac{\partial R_k^{(\varphi)}}{\partial t}\right)^{\alpha'}_{\ \
\alpha}\frac{\partial\varphi^\alpha_k}{\partial\psi^\beta}
\nonumber\\
&&+\frac{1}{2}\left(\hat\Gamma_k^{(2)}\right)^{-1\beta}_{\  \ \ \beta'}
\frac{\partial\varphi^*_{k\alpha'}}{\partial\psi^*_{\beta'}}
\left(\frac{\partial R_k^{(\varphi)}}
{\partial t}\right)^{\alpha'}_{\ \alpha}
\frac{\partial\varphi_k^{\ \alpha}}
{\partial\psi^\beta}\nonumber\\
&&+\frac{\partial\tilde\Gamma_k[\hat\varphi,\psi]}
{\partial\hat\varphi^\alpha}
\frac{\partial\varphi^\alpha_k[\psi]}{\partial
t}-\frac{\partial\Delta_k^{(\varphi)}S[\hat\varphi,\psi]}{\partial\hat
\varphi^\alpha}\frac{\partial\varphi^\alpha_k[\psi]}{\partial t}\ear
The elements of the covariant second functional derivative
$\hat\Gamma_k^{(2)}$
read
\bear\label{A.9}
&&\left(\hat\Gamma_k^{(2)}\right)^\alpha_{\ \alpha'}=\frac{\partial^2
\tilde\Gamma_k[\hat\varphi,\psi]}
{\partial\hat\varphi_\alpha^*\partial\hat\varphi^{\alpha'}}\nonumber\\
&&\left(\hat\Gamma_k^{(2)}\right)^\alpha_{\ \beta}=\frac{\partial^2
\tilde\Gamma_k[\hat\varphi,\psi]}
{\partial\hat\varphi_\alpha^*\partial\psi^{\beta}},\quad
\left(\hat\Gamma_k^{(2)}\right)^\beta_{\ \alpha}=\frac{\partial^2
\tilde\Gamma_k[\hat\varphi,\psi]}
{\partial\psi^*_\beta\partial\hat\varphi^{\alpha}}\nonumber\\
&&\left(\hat\Gamma_k^{(2)}\right)^\beta_{\ \beta'}=\frac{\partial^2
\tilde\Gamma_k[\hat\varphi,\psi]}
{\partial\psi_\beta^*\partial\psi^{\beta'}}
-\frac{\partial^2\varphi^\alpha_k}{\partial\psi^*_\beta\partial
\psi^{\beta'}}
\frac{\partial\tilde\Gamma_k[\hat\varphi,\psi]}{\partial
\hat\varphi^\alpha}\ear
We note that eqs. (\ref{A.8}),(\ref{A.9}) hold for arbitrary
$\varphi_k[\psi]$ in the transformation (\ref{A.7}). If we use
in addition the definition
$\varphi_k[\psi]$ as a solution of the field equation (\ref{2.13}),
we have the properties
\be\label{A.10}
\Gamma_k[\hat\varphi=0,\psi]=\Gamma_k[\psi]-\Delta^{(\varphi)}_k
S[\varphi_k[\psi]]\ee
\be\label{A.11}
\frac{\partial\tilde\Gamma_k}{\partial\hat\varphi^\alpha}
_{|\hat\varphi=0}=0\ee
In the evolution equation for $\frac{\partial}{\partial t}\Gamma_k[\psi]=
\frac{\partial}{\partial t}\Gamma_k[0,\psi]+\frac{\partial}{\partial t}
\Delta_k^{(\varphi)}S[\varphi_k]$ the covariant second
functional derivative $\hat\Gamma_k^{(2)}$ reduces to the simple second
functional derivative of $\tilde \Gamma_k$ with respect to $\hat\varphi$
and
$\psi$. Also $\left(\hat\Gamma_k^{(2)}\right)^\alpha_{\ \beta}$ and
$\left(\hat\Gamma_k^{(2)}\right)^\beta_{\ \alpha}$
vanish and $\hat\Gamma_k^{(2)}$ is therefore block-diagonal in the
$\varphi$
and $\psi$ components, implying in turn $\left(\hat\Gamma_k^{(2)}
\right)^{-1\alpha}_{\ \ \  \beta}=\left(\hat\Gamma_k^{(2)}\right)
^{-1\beta}_{\ \ \ \alpha}=0$. The $\psi-\psi$ components of the second
functional derivative $\left(\hat\Gamma_k^{(2)}\right)_{\ \beta}^{\beta'}$
are simply related to the second functional derivative of $\Gamma_k[\psi]$
\be\label{A.12}
\left(\hat\Gamma_k^{(2)}\right)_{\
\beta}^{\beta'}=\left(\hat\Gamma_k^{(2)}
[\psi]+R^{(\psi)}_k\right)_{\ \beta}^{\beta'}\ee

\section*{Appendix B: Alternative formulation}
\renewcommand{\theequation}{B.\arabic{equation}}
\setcounter{equation}{0}

Although suggestive, the definition of $\Gamma_k[\psi]$ by eqs.
(\ref{2.11})
and (\ref{2.13}) is not the only possibility for an effective action for
$\psi$. As an alternative we propose to define $\bar\varphi_k[\psi]$ as
the
solution of the field equation derived from $\Gamma_k[\varphi,\psi]$
instead
of $\tilde\Gamma_k[\varphi,\psi]$, i.e. we replace (\ref{2.12}) and
(\ref{2.13}) by
\be\label{B.1}
\frac{\partial\Gamma_k[\varphi,\psi]}{\partial\varphi^\alpha}
_{|\bar\varphi_k[\psi]}=0\ee
\be\label{B.2}
\bar\Gamma_k[\psi]=\Gamma_k[\bar\varphi_k[\psi],\psi]\ee
We briefly  discuss the modifications of our formalism and compare the
properties of $\Gamma_k[\psi]$ and $\bar\Gamma_k[\psi]$ at the end of
this section. The functional $\bar\Gamma_k[\psi]$ is not related to the
Legendre transform of $W_k[K]$ (\ref{2.10}), but
rather to $W_k[J,K]$ with $J$ given by
$R_k^{(\varphi)}\bar\varphi_k[\psi]$.
In the limit $k\to0$ the difference $\Delta_kS$ between $\tilde\Gamma_k[
\varphi,\psi]$ and $\Gamma_k[\varphi,\psi]$ vanishes. We conclude that
$\bar\Gamma_0[\psi]$ and $\Gamma_0[\psi]$ coincide
\be\label{B.3}
\Gamma_0[\psi]=\bar\Gamma_0[\psi]\ee
and both define the generating functional for the 1PI Green functions
for $\psi$. Inserting the classical solution $\bar\varphi_k[\psi]$ instead
of $\varphi_k[\psi]$ in the evolution equation (A.8), (A.9) leads to
three modifications: First, the last two terms in eq. (A.8) are now
absent.
Second, the derivatives
\bear\label{B.4}
&&\frac{\partial\tilde\Gamma_k}{\partial
\hat\varphi^*_\alpha}[\hat\varphi=0,
\psi]=(R_k^{(\varphi)})^\alpha_{\ \alpha'}\varphi^{\alpha'}_k[\psi]
\nonumber\\
&&\frac{\partial\tilde\Gamma_k}{\partial
\hat\varphi^\alpha}[\hat\varphi=0,
\psi]=\varphi^*_{k\alpha'}[\psi](R_k^{(\varphi)})^{\alpha'}_{\ \ \alpha}
\ear
induce an additional contribution in $(\hat\Gamma_k^{(2)})^\beta_{\
\beta'}$
which now reads
\bear\label{B.5}
&&(\hat\Gamma_k^{(2)})^\beta_{\ \beta'}=(\bar\Gamma_k^{(2)}[\psi]+\check
{R}_k[\psi])^\beta_{\ \beta'}\nonumber\\
&&(\check{R}_k[\psi])^{\beta'}_{\ \ \beta}=(R_k^{(\psi)})^{\beta'}_{\ \
\beta}+\frac{\partial\varphi^*_{k\alpha'}}{\partial\psi^*_{\beta'}}
(R_k^{(\varphi)})^{\alpha'}_{\ \ \alpha}\frac{\partial\varphi^\alpha_k}
{\partial\psi^\beta}\ear
Third, the off-diagonal elements of $\hat\Gamma_k^{(2)}$ do not vanish
anymore
\bear\label{B.6}
&&(\hat\Gamma_k^{(2)})^\alpha_{\ \beta}=(R_k^{(\varphi)})^\alpha_{\
\alpha'}
\frac{\partial\varphi^{\alpha'}_k[\psi]}{\partial\psi^\beta}\nonumber\\
&&(\hat\Gamma_k^{(2)})^\beta_{\ \alpha}=
\frac{\partial\varphi^*_{k\alpha'}[\psi]}{\partial\psi^*_\beta}
(R_k^{(\varphi)})^{\alpha'}_{\ \ \alpha}\ear

The role of these modifications can best be understood if we write
formally
the original evolution equation (2.9) as
\be\label{B.7}
\frac{\partial}{\partial
t}\Gamma_k[\varphi,\psi]=\frac{1}{2}\tilde\partial
_t\ln\det(\Gamma_k^{(2)}+R_k)\ee
where $\tilde\partial_t$ acts only on $k$. We can express
$\Gamma_k$ in terms of the new variables $\hat\sigma$ (A.2) and write
\be\label{B.8}
\frac{\partial}{\partial
t}\Gamma_k[\hat\sigma]=\frac{1}{2}\tilde\partial_t
\ln\det\left(\frac{\partial\sigma^*_\gamma}
{\partial\hat\sigma^*_{\gamma'}}
\left(\Gamma_k^{(2)}+R_k\right)^\gamma_{\
\delta}\frac{\partial\sigma^\delta}
{\partial\hat\sigma^{\delta'}}\right)\ee
provided the transformation $\hat\sigma
(\sigma)$ does not explicitly
or implicitly involve $R_k$. This is realized
for the definition of
$\bar\varphi_k$ by (B.1).
Inserting $\hat\varphi=0$ and using eq. (B.1), one obtains
\be\label{B.9}
\frac{\partial}{\partial t}\bar\Gamma_k[\psi]=\frac{1}{2}Tr\left\{
\frac{\partial\check{R}_k[\psi]}{\partial t}(\check{\Gamma}_k^{(2)}
[\psi]+\check{R}_k[\psi])^{-1}
\right\}\ee
in agreement with (A.8), (B.5), (B.6). Here
$\check{\Gamma}_k^{(2)}$ and $\check{R}
_k$ are matrices with $\alpha$ and $\beta$ indices, and
$\check{\Gamma}_k^{(2)}$
is block diagonal
\bear\label{B.10}
(\check{\Gamma}_k^{(2)})^{\beta'}_{\ \ \beta}&=&
\frac{\partial^2\bar\Gamma_k[\psi]}{\partial\psi_{\beta'}^*
\partial\psi^\beta}\nonumber\\
(\check{\Gamma}_k^{(2)})^{\alpha'}_{\ \ \alpha}&=&(\Gamma_k^{(2)}
[\varphi,\psi])^{\alpha'}_{\ \ \alpha_{|
\varphi=\bar\varphi_k[\psi]}}\ear
The matrix
\be\label{B.11}
(\check{R}_k)^{\gamma'}_{\ \ \delta'}=\frac{\partial\sigma_\gamma}
{\partial\hat\sigma^*_{\gamma'}}
(R_k)^\gamma_{\ \delta}\frac{\partial\sigma^\delta}{\partial
\hat\sigma^{\delta'}}\ee
has diagonal and off-diagonal elements as given by
\bear\label{B.12}
&&(\check{R}_k)^{\beta'}_{\ \ \beta}=(R^{(\psi)}_k)^{\beta'}_
{\ \ \beta}+\frac
{\partial\bar\varphi_{k\alpha'}^*}{\partial\psi^*_{\beta'}}
(R^{(\varphi)}_k)^{\alpha'}_{\ \ \alpha}\frac
{\partial\bar\varphi_k^\alpha}{\partial\psi^\beta}\nonumber\\
&&(\check{R}_k)^{\alpha'}_{\ \ \alpha}=(R^{(\varphi)}_k)^
{\alpha'}_{\ \ \alpha}
\nonumber\\
&&(\check{R}_k)^{\alpha'}_{\ \ \beta}=(R^{(\varphi)}_k)^
{\alpha'}_{\ \ \alpha}
\frac{\partial\bar\varphi_k^\alpha}{\partial\psi^\beta}\nonumber\\
&&(\check{R}_k)^{\beta'}_{\ \ \alpha}=\frac{\partial\bar\varphi^*
_{k \alpha'}}{\partial\psi^*_{\beta'}}
(R^{(\varphi)}_k)^{\alpha'}_{\ \ \alpha}\ear
Similar as to the evolution equation for $\Gamma_k[\psi]$ (\ref{2.17})
we need information on the $\psi$-dependence of $\bar\varphi_k$
and on $\tilde
\Gamma_k^{(2)}[\bar\varphi_k,\psi]$ in order to specify the
evolution equation (B.9) for $\bar\Gamma_k[\psi]$.
It is convenient to split $\check{\Gamma}_k^{(2)}+\check{R}_k$ into
a block-diagonal and off-diagonal piece
\bear\label{B.13}
&&\check{\Gamma}_k^{(2)}+\check{R}_k=D_k+\Delta_k\nonumber\\
&&(\Delta_k)^{\alpha'}_{\ \ \beta}=(R_k^{(\varphi)})^{\alpha'}
_{\ \ \alpha}
\frac{\partial\varphi^\alpha_k}{\partial\psi_\beta}\nonumber\\
&&(\Delta_k)^{\beta'}_{\ \ \alpha}=\frac{\partial\varphi^*_{k\alpha'}}
{\partial
\psi^*_{\beta'}}(R_k^{(\varphi)})^{\alpha'}_{\ \ \alpha}\ear
and to expand in $\Delta$, using  iteratively
the identity
\be\label{B.14}
(D+\Delta)^{-1}=D^{-1}-D^{-1}\Delta(D+\Delta)^{-1}\ee
Even powers of $\Delta$ contribute only to the block-diagonal piece of
$(D+\Delta)^{-1}$, i.e.
\bear\label{B.15}
H^{-1}&=&D^{-1}+D^{-1}\Delta D^{-1}\Delta D^{-1}+D^{-1}\Delta
D^{-1}\Delta D^{-1}
\Delta D^{-1}\Delta D^{-1}+...\nonumber\\
&=&(D-\Delta D^{-1}\Delta)^{-1}\ear
whereas odd powers of $\Delta$ give an off-diagonal contribution.

In summary, we can write the evolution equation (B.9) in
the explicit form
\bear\label{B.16}
&&\frac{\partial}{\partial t}\bar\Gamma_k[\psi]=\frac{1}{2}
\frac{\partial}
{\partial t}(R_k^{(\varphi)})^\alpha_{\ \alpha'}(H^{-1})^
{\alpha'}_{\ \ \alpha}
\nonumber\\
&&+\frac{1}{2}\frac{\partial}{\partial t}(\check{R}_k)
^\beta_{\ \beta'}(H^{-1})^{\beta'}_{\ \ \beta}\nonumber\\
&&-\frac{1}{2}\frac{\partial}{\partial t}(\check{R}_k)
^\alpha_{\ \beta'}(D^{-1})^{\beta'}_{\ \ \beta}(\check{R}_k)
^\beta_{\ \alpha'}(H^{-1})^{\alpha'}_{\ \ \alpha}\nonumber\\
&&-\frac{1}{2}\frac{\partial}{\partial t}(\check{R}_k)
^\beta_{\ \alpha'}(D^{-1})^{\alpha'}_{\ \ \alpha}(\check{R}_k)
^\alpha_{\ \beta'}(H^{-1})^{\beta'}_{\ \ \beta}\ear
with
\bear\label{B.17}
&&H^\alpha_{\ \alpha'}=D^\alpha_{\ \alpha'}-(\check R_k)^\alpha_
{\ \beta'}(D^{-1})
^{\beta'}_{\ \ \beta}(\check{R}_k)^\beta_{\ \alpha'}\nonumber\\
&&H^\beta_{\ \beta'}=D^\beta_{\ \beta'}-(\check R_k)^\beta_
{\ \alpha'}(D^{-1})
^{\alpha'}_{\ \ \alpha}(\check{R}_k)^\alpha_{\ \beta'}\ear
and
\bear\label{B.18}
&&D^\alpha_{\ \alpha'}=(\Gamma_k^{(2)}[\bar\varphi_k[\psi],\psi])
^\alpha_{\
\alpha'}+(R_k^{(\varphi)})^\alpha_{\ \alpha'}\nonumber\\
&&D^\beta_{\ \beta'}=(\bar\Gamma_k^{(2)}[\psi])^\beta_{\
\beta'}+(\check{R}_k)^\beta_{\ \beta'}\ear

The main difference with the formulation in sect. 2 is the absence
of an effective infrared cutoff in the classical solution (\ref{B.1})
if $\varphi$ is a massless field. This solution and therefore the flow
equation (\ref{B.16}) typically involves then nonlocalities of the form
$1/q^2$.
The effective action $\bar\Gamma_k$ (B.2) equals in lowest (classical)
approximation the classical approximation to $\bar\Gamma_0$ and the
scale $k$ appears only in the contributions from quantum fluctuations. In
contrast to $\Gamma_k$ non-local $n$-point functions
are therefore already present for
$k>0$. At first sight this may seem as an advantage of the use
of $\bar\Gamma_k$ since in the classical approximation the nonlocalities
in $\Gamma_k$ have to build up in the course of the scale evolution
instead of being included from the outset. Going beyond the classical
approximation there is, however, a considerable price to pay: First, the
flow equation (\ref{B.16}) is more complicated than (\ref{2.17}). Also
at every scale $k$ one has to deal with the nonlocality of the classical
solution instead of the essentially local behaviour of $\Gamma_k$. For
these reasons we will use in the present paper the version (\ref{2.17})
of the flow equation.

\section*{Appendix C: Field transformation for fermionic models}
\setcounter{equation}{0}
\renewcommand{\theequation}{C.\arabic{equation}}
The formalism of Appendix A can easily be generalized for fermions,
with a little care for minus signs from the anticommutation of
Grassmann variables. The generalization of (\ref{A.1}) reads for the
special case (\ref{A.7}), (\ref{A.10}), (\ref{A.11})
\be\label{C.1}
\frac{\partial}{\partial t}\Gamma_k[\psi]=\frac{1}{2}STr\left
\lbrace\left(\tilde \Gamma_k^{(2)}\right)^{-1}\frac{\partial R_k}
{\partial t}
\right\rbrace+\frac{\partial}{\partial t}\Delta_k^{(\varphi)}S_{|_{\varphi
_k}}\ee
Here the supertrace $STr$ contains an additional minus sign for the
fermionic part of the trace. Using $STr (AB)=STr (BA)$ we can write
(with the notations of appendix A)
\be\label{C.2}
STr\left\lbrace\left(\tilde\Gamma_k^{(2)}\right)^{-1}\frac{\partial R_k}
{\partial t}\right\rbrace=STr\left\lbrace C^{-1}\frac{\partial\hat R_k}
{\partial t}\right\rbrace\ee
with
\be\label{C.3}
\frac{\partial}{\partial t}
\hat R^\alpha_{\ \beta}=\tilde\kappa
\frac{\partial\sigma_{\alpha'}^*}{\partial\hat
\sigma_\alpha^*}\left(\frac{\partial R_k}{\partial t}\right)^{\alpha'}_{\
\beta'}\frac{\partial\sigma^{\beta'}}{\partial\hat\sigma^\beta}\ee
and
\be\label{C.4}
C^\alpha_{\ \beta}=\tilde\kappa
\frac{\partial\sigma_{\alpha'}^*}{\partial\hat
\sigma_\alpha^*}\left(\tilde\Gamma_k^{(2)}\right)^{\alpha'}_{\
\beta'}\frac{\partial\sigma^{\beta'}}{\partial\hat\sigma^\beta}\ee
Here we have introduced for later convenience a factor $\tilde\kappa=
i^N$ with $N$ the number of mixed bosonic/fermionic derivatives of the
type $\partial\hat\varphi^*/\partial\bar\psi$ or $\partial\varphi/\partial
\hat\psi$ etc.
Employing (A.11) one has
\be\label{C.5}
\left(\tilde\Gamma_k^{(2)}\right)^{\alpha'}_{\ \beta'}
=\kappa\frac{\partial\hat\sigma_{\gamma}^*}{\partial
\sigma_{\alpha'}^*}\left(\hat\Gamma_k^{(2)}\right)^{\gamma}_{\
\delta}\frac{\partial\hat\sigma^{\delta}}{\partial\sigma^{\beta'}}\ee
where
\be\label{C.6}
\left(\hat\Gamma_k^{(2)}\right)^\gamma_{\ \delta}=\frac{\partial^2\tilde
\Gamma_k}{\partial\hat\sigma^\delta\partial\hat\sigma^*_\gamma}
_{|\hat\varphi=0}\ee
is block diagonal in the bosonic and fermionic subspaces and therefore
a bosonic quantity. The factor $\kappa$ arises from the commutation
of $(\partial\hat\sigma_\gamma^*/\partial\sigma_\alpha^*)$ with
$(\partial/\partial\sigma^\beta)$ and obeys
\be\label{C.7}
\kappa=\left\lbrace\begin{array}{lll}-1& {\rm for}&
\hat\sigma^*=\varphi^*,
\hat\sigma=\varphi,
\sigma^*=\bar\psi,\sigma=\psi\\ 1&{\rm else}& \\ \end{array}\right.
\ee
where $\varphi$ is bosonic and $\psi$ fermionic. In consequence
one has $\kappa=\tilde\kappa$, and we observe
$C=\hat\Gamma_k^{(2)}$.  This yields finally
\be\label{C.8}
STr\left\lbrace\left(\tilde\Gamma_k^{(2)}\right)^{-1}
\frac{\partial R_k}{\partial t}\right\rbrace=STr\left\lbrace
\left(\hat\Gamma_k^{(2)}\right)^{-1}
\frac{\partial \hat R_k}{\partial t}\right\rbrace\ee
where the bosonic part of $\hat R_k$ is simply given by $R_k^{(\varphi)}$
whereas the fermionic part reads
\be\label{C.9}
\hat R_k^{(\psi)}=R_k^{(\psi)}-\frac{\partial\varphi^*_k}
{\partial \bar\psi}
R_k^{(\varphi)}\frac{\partial\varphi_k}{\partial\psi}\ee
The present formulation was adapted for Majorana spinors where $\psi$
and $\bar\psi$ are not independent. For Dirac spinors the factor $1/2$
in front of the fermionic trace is absent. This amounts to a
multiplication
of $R_k^{(\psi)}$ by two. Furthermore, $\psi$ and $\bar\psi$ are now
independent and (\ref{C.9}) must be supplemented by a second term where
the role of $\psi$ and $\bar\psi$ is exchanged. These two identical
contributions
also multiply effectively the second term in (\ref{C.9})
by a factor of two.
The overall modification for Dirac spinors is a multiplication of
$\hat R^{(\psi)}$ by a factor two.

\section*{Appendix D: Classical field equations and propagators}
\setcounter{equation}{0}
\renewcommand{\theequation}{D.\arabic{equation}}
In this appendix we derive the field equations and the gauge field
propagator
from the effective average action $\Gamma_k[\psi,A,\bar A=0]$
in the truncation given by (\ref{4.1}). We start with the gauge-invariant
kinetic term
\be\label{D.1}
\Gamma^{(F)}_{kin}=\frac{\tilde Z_F}{4}\int d^d x F_z^{\mu\nu}
F_{\mu\nu}^z\ee
\be\label{D.2}
F^z_{\mu\nu}=\partial_\mu A^z_\nu-\partial_\nu A^z_\mu+\tilde g f_{wy}^
{\phantom{wy}z}A_\mu^w A_\nu^y
\quad,\quad\tilde g= g_k\tilde Z_{F,k}^{\frac{1}{2}}
\ee
and write it as a sum of terms quadratic, cubic and quartic in $A$
\be\label{D.3}
\Gamma^{(F)}_{kin}=\Gamma^{(F)}_{kin,2}+\Gamma^{(F)}_{kin,3}+\Gamma^
{(F)}_{kin,4}\ee
with
\bear
\Gamma^{(F)}_{kin,2}&=&\frac{\tilde Z_F}{2}\int d^dx A^\nu_
z(\partial_\nu\partial^\mu-\partial^2\delta^\mu_\nu)A^z_\mu\label{D.4}\\
\Gamma^{(F)}_{kin,3}&=&\tilde g\frac{\tilde Z_F}{2}
f_{wy}^{\phantom{wy}z}
\int d^d  x A^w_\mu A^y_\nu(\partial^\mu A^\nu_z-\partial^\nu A^\mu_z)
\label{D.5}\\
\Gamma^{(F)}_{kin,4}&=&\tilde g^2\frac{\tilde Z_F}{4} C^{yz}_{wx}\int
d^dx A^w_\mu A^\mu_y A^x_\nu A^\nu_z\label{D.6}
\ear
and
\be\label{D.7}
C^{yz}_{wx}=f_{wx}^{\phantom{wx} u} f^{yz}_{\phantom{yz}u}=-(T^u)_
{wx}(T_u)^{yz}
\ee
(For the last equations we use $f_{wx}^{\phantom{wx}u}=f^u_
{\ wx}=i(T^u)_{wx})$. For our  purposes it is convenient to work in
momentum space where
\bear
\Gamma^{(F)}_{kin,2}&=&\frac{\tilde Z_F}{2}\int\frac{d^dq}{(2\pi)^d}
A^\nu_z(-q)A^z_\mu(q)(q^2\delta^\nu_\mu-q^\nu q_\mu)\label{D.8}\\
\Gamma^{(F)}_{kin,3}&=&i\tilde g\frac{\tilde Z_F}{2}
f_{wy}^{\phantom{wy}z}
\int\frac{d^dp_1}{(2\pi)^d}\frac{d^dp_2}{(2\pi)^d}\frac{d^dp_3}
{(2\pi)^d}(2\pi)^d
\delta(p_1-p_2-p_3)\nonumber\\
&&(p_1^\mu A^\nu_z(-p_1)-p_1^\nu A^\mu_z(-p_1))A^w_\mu(p_2)A^y_
\nu(p_3)\label{D.9}\\
\Gamma^{(F)}_{kin,4}&=&\tilde g^2\frac{\tilde Z_F}{4}C^{yz}_{wx}\int
\frac{d^dp_1}{(2\pi)^d}\cdots \frac{d^dp_4}{(2\pi)^d}(2\pi)^d\delta
(p_1+p_2-p_3-p_4)\nonumber\\
&&A^w_\mu(-p_1) A^\mu_y(p_3) A^x_\nu(-p_2)A^\nu_z(p_4).\label{D.10}
\ear
Similarly the gauge invariant fermionic kinetic term reads
\bear\label{D.11}
\Gamma^{(\psi)}_{kin}&=&Z_\psi\int\frac{d^d q}{(2\pi)^d}\bar\psi^i_a(q)
\gamma^\mu q_\mu\psi^a_i(q)\nonumber\\
&&+\tilde g Z_\psi\int\frac{d^dp}{(2\pi)^d}\frac{d^dp'}{(2\pi)^d}
\frac{d^dq}{(2\pi)
^d}(2\pi)^d\delta(p-p'+q)\nonumber\\
&&\ \bar\psi^i_a(p)\gamma^\mu(T_z)_i^{\ j}\psi^a_j(p')A^z_\mu(-q)
\ear
The field equation obtains as
\bear\label{D.12}
&&\frac{\delta}{\delta
A^z_\mu(-q)}\left(\Gamma_{kin}^{(\psi)}+\Gamma_{kin}
^{(F)}+\Gamma_k^{gauge}
+\Delta_kS^{(A)}\right)_{|\bar A=0}=\nonumber\\
&&\tilde g Z_\psi\int\frac{d^dp}{(2\pi)^d}\bar\psi^i_a(p)
\gamma^\mu(T_z)_i^{\ j}\psi^a_j(p+q)+\tilde Z_FP(q)A^\mu_z(q)
\nonumber\\
&&+i\tilde g\tilde Z_Ff_{zy}^{\ \ w}\int\frac{d^dp}{(2\pi)^d}
A^y_\nu(q-p)\left\lbrace(q^\nu+p^\nu)A^\mu_w(p)
-p^\mu A^\nu_w(p)\right\rbrace\nonumber\\
&&+\tilde g^2\frac{\tilde Z_F}{2}\tilde C_{zx yw}\int\frac{d^dp}{(2\pi)^d}
\frac{d^dp'}{(2\pi)^d}A^{y\mu}(q+p-p')A^x_\nu(-p)A^{w\nu}(p')=0\ear
where we use the invariant tensor
\be\label{D.13}
\tilde C_{zxyw}=f_{zx}^{\ \ u}f_{ywu}+f_{zw}^{\ \ u}f_{yxu}\ee
For an expansion in powers of $\bar\psi\psi$ we
can solve the field equation (\ref{D.12}) iteratively
\begin{displaymath}
(A_k(q))^\mu_z=(A_k^{(0)}(q))^\mu_z+(A_k^{(1)}(q))^\mu_z+...
\end{displaymath}
with the result
\be\label{D.14}
(A_k^{(0)}(q))^\mu_z=-\tilde g\frac{Z_\psi}{\tilde
Z_F}P^{-1}(q)\int\frac{d^dp}{(2\pi)^d}\bar\psi_a^i(p)\gamma^
\mu(T_z)_i^{\ j}
\psi^a_j(p+q)\ee
\bear\label{D.15}
&&(A_k^{(1)}(q))^\mu_z=-i\tilde gP^{-1}(q)f_{zy}^{\ \ w}\int\frac{d^dp}
{(2\pi)^d}(A_k^{(0)}(q-p))^y_\nu\nonumber\\
&&\left\lbrace\left(q^\nu+p^\nu\right)\left(A_k^{(0)}(p)\right)^\mu_w-
p^\mu\left(A^{(0)}_k(p)\right)^\nu_w\right\rbrace\nonumber\\
&&=i\tilde g^3\frac{Z_\psi^2}{\tilde Z_F^{\ 2}}P^{-1}(q)
f_{zy}^{\ \ w}\int\frac{
d^dp_1}{(2\pi)^d}...\frac{d^dp_4}{(2\pi)^d}(2\pi)^d\delta(q-p_1-p_2+
p_3+p_4)
\nonumber\\
&&P^{-1}(p_1-p_3)P^{-1}(p_2-p_4)\Bigl[(p_2-p_4)^\mu\{\bar\psi(-p_1)
\gamma_\nu T^y\psi(-p_3)\}
\{\bar\psi(p_4)\gamma^\nu T_w\psi(p_2)\}\nonumber\\
&&+(p_1-p_3-2q)^\nu
\{\bar\psi(-p_1)\gamma_\nu T^y\psi(-p_3)\}\{\bar\psi(p_4)
\gamma^\mu T_w\psi(p_2)\}\Bigr]\ear

The effective inverse gauge field propagator $(\Gamma_k^{(2)})
^{y\mu}_{\nu z}
(q',q)+R^{(A)}_k(q)\delta_z^y\delta^\mu_\nu(2\pi)^d\delta(q-q')$
is related to the second functional derivatives
\bear\label{D.16}
&&\frac{\delta^2}{\delta A^z_\mu(-q)\delta A^\nu_y(q')}
\left(\Gamma_{kin,2}^{(F)}
+\Gamma^{gauge}_k+\Delta_kS^{(A)}\right)_{|\bar A=0}\nonumber\\
&&=\tilde Z_FP(q)\delta^y_z\delta_\nu^\mu(2\pi)^d\delta(q-q')\ear
\bear\label{D.17}
&&\frac{\delta^2\Gamma_{kin,3}^{(F)}}{\delta A^z_\mu(-q)\delta A^
\nu_y(q')}=i\tilde g\tilde Z_Ff_z^{\ yw}\left\lbrace
(2q-q')_\nu A^\mu_w(q-q')\right.\nonumber\\
&&\left.+(2q'-q)^\mu A_{w\nu}(q-q')-(q+q')_\sigma A^\sigma_w(q-q')
\delta^\mu_\nu\right\rbrace\ear
\bear\label{D.18}
&&\frac{\delta^2\Gamma_{kin,4}^{(F)}}{\delta A^z_\mu(-q)
\delta A^\nu_y(q')}
=\tilde g^2\frac{\tilde Z_F}{2}\int\frac{d^dp}{(2\pi)^d}\cdot\nonumber\\
&&\left\lbrace\tilde C_{zx\ w}^{\ \ y}\delta^\mu_\nu A^x_
\sigma(q-q'+p)A^{w\sigma}
(-p)+2\tilde C_{z\ xw}^{\ y} A^{x\mu}(q-q'+p)A^w_\nu(-p)\right\rbrace
\ear
It is now straightforward to expand the effective
propagator $(\Gamma_k^{(2)}
+R_k^{(A)})^{-1}$ in powers of $A$:
\bear\label{D.19}
&&\left[\left(\Gamma_k^{(2)}+R_k^{(A)}\right)^{-1}\right]^{y\mu}_
{\nu z}(q',q)=\tilde Z_F^{-1}P(q)^{-1}\delta^y_z
\delta^\mu_\nu(2\pi)^d\delta(q-q')
\nonumber\\
&&-i\tilde g\tilde Z_F^{-1}P(q')^{-1}P(q)^{-1}f_z^{\ yw}.\nonumber\\
&&\lbrace(2q-q')_\nu A^\mu_w(q-q')+(2q'-q)^\mu A_{w\nu}(q-q')-
(q+q')_\sigma A^\sigma_w(q-q')\delta^\mu_\nu\rbrace\nonumber\\
&&-\frac{1}{2}\tilde g^2\tilde Z_F^{-1}P(q')^{-1}P(q)^{-1}
\int\frac{d^dp}{(2\pi)^d}\cdot\\
&&\left\lbrace\tilde C_{zx\ w}^{\ \ y}\delta^\mu_\nu
A^x_\sigma(q-q'+p)A^{w\sigma}(-p)+2\tilde C^{\ y}_{z\ xw}
A^{x\mu}(q-q'+p)A^w_\nu(-p)\right\rbrace\nonumber\\
&&-\tilde g^2\tilde Z_F^{-1}P(q')^{-1}P(q)^{-1}\int\frac{d^dp}
{(2\pi)^d}P(p)^{-1}
f_{y'}^{\ yw}f_z^{\ y'w'}.\nonumber\\
&&\left\lbrace(2p-q')_\nu A^{\nu'}_w(p-q')+(2q'-p)^{\nu'}
A_{w\nu}(p-q')-(p+q')_\sigma A^\sigma_w(p-q')
\delta^{\nu'}_\nu\right\rbrace.\nonumber\\
&&\left\lbrace(2q-p)_{\nu'} A^{\mu}_{w'}(q-p)+(2p-q)^{\mu}A_{w'\nu'}(q-p)
-(q+p)_\tau A^\tau_{w'}(q-p)\delta^{\mu}_{\nu'}\right\rbrace+0(A^3)
\nonumber\ear
Eqs. (D.14), (D.15), and (D.19) define the quantities
needed for $\gamma_A$
and $\gamma_{A\psi}$ in sect. 4.
For easy reference, we also give the vertices
\bear\label{D.20}
&&\frac{\delta^3\Gamma_{kin,3}^{(F)}}{\delta A^{\mu z}(q_1)\delta A
^{\nu y}(q_2)\delta A^{\sigma w}(q_3)}=i\tilde g
\tilde Z_Ff_{zyw}(2\pi)^d\delta(q_1+q_2+q_3)\nonumber\\
&&\left\lbrace(q_1-q_2)_\sigma\delta_{\mu\nu}+
(q_2-q_3)_\mu\delta_{\nu\sigma}
+(q_3-q_1)_\nu\delta_{\mu\sigma}\right\rbrace\ear
\bear\label{D.21}
&&\frac{\delta^3\Gamma_{kin,4}^{(F)}}{\delta A_\mu^z(-q)\delta A
^\nu_y(q')\delta A_\sigma^w(-p)}=\tilde g^2 \tilde Z_F
\left\lbrace\tilde C_{zw\ x}^{\ \ y}\delta^\mu_\nu A^{x\sigma}(p+q-q')
\right.\nonumber\\
&&+\tilde C_{z\ wx}^{\ y}\delta^{\mu\sigma}A^x_\nu(p+q-q')
\left.+\tilde C_{z\ xw}^{\ y}\delta_\nu^\sigma
A^{x\mu}(p+q-q')\right\rbrace
\ear
\bear\label{D.22}
&&\frac{\delta^4\Gamma_{kin,4}^{(F)}}{\delta A_\mu^z(-q)\delta A
^\nu_y(q')\delta A_\sigma^w(-p)\delta A^\tau_x(p')}=\tilde g^2
\tilde Z_F(2\pi)^d\delta(q+p-q'-p')\nonumber\\
&&\left\lbrace\tilde C_{zw}^{\ \ yx}\delta^\mu_\nu
\delta^\sigma_\tau+
\tilde C_{z\ w}^{\ y\ x}\delta^{\mu\sigma}\delta_{\nu\tau}+
\tilde C_{z\ \ w}^{\ yx}\delta_\nu^\sigma \delta^\mu_\tau\right\rbrace
\ear

\section*{Appendix E: Expansion of the effective action and propagator in
fermion bilinears}
\setcounter{equation}{0}
\renewcommand{\theequation}{E.\arabic{equation}}
The effective action for quarks $\Gamma_k[\psi]$ contains an equal number
of $\psi$ and $\bar\psi$ fields as a consequence of baryon number
conservation. It can therefore be expanded in terms of 1PI functions for
an even number of fields. Denoting by $a.., i.., \gamma..(\dot\gamma..)$
the
flavour, colour, and spinor indices of the Dirac spinors $\psi(\bar\psi)$,
one has in a momentum basis
\bear\label{E.1}
&&\Gamma_k[\psi]=\Gamma_k^{(0)}[0]
+\int\frac{d^dq}{(2\pi)^d}\frac{d^dq'}{(2\pi)^d}\bar\psi^{i\dot\gamma}_{\
a}(q)\psi^b_{j\gamma}(q')\left(\Gamma_k^{(2)}[0]\right)
^{aj\gamma}_{bi\dot\gamma}(q,q')\nonumber\\
&&+\frac{1}{4}\int\frac{d^dp_1}{(2\pi)^d}...
\frac{d^dp_4}{(2\pi)^d}\bar\psi^{i\dot\gamma}_a(-p_1)
\psi^b_{j\gamma}(p_2)\bar\psi^{k\dot\delta}_c(p_4)\psi^d_{l\delta}(-p_3)
\nonumber\\
&&\left(\Gamma_k^{(4)}[0]\right)^{acjl\gamma\delta}_{bdik\dot\gamma
\dot\delta}
(p_1,p_2,p_3,p_4)+0[(\bar\psi\psi)^3]\ear
As a consequence of momentum conservation and global colour symmetry
we write
\be\label{E.2}
\left(\Gamma^{(2)}_k[0]\right)^{aj\gamma}_{bi\dot\gamma}(q,q')
=\left(G^{(2)}
\right)^{a\gamma}_{b\dot\gamma}(q)\delta^j_i(2\pi)^d\delta(q-q')\ee
\be\label{E.3}
\left(\Gamma^{(4)}_k[0]\right)^{acjl\gamma\delta}
_{bdik\dot\gamma\dot\delta}
(p_1,p_2,p_3,p_4)=\left(G^{(4)}
\right)^{acjl\gamma\delta}_{bdik\dot\gamma\dot\delta}
(p_1,p_2,p_3,p_4)
(2\pi)^d\delta(p_1+p_2-p_3-p_4)\ee
with
\bear\label{E.4}
(G^{(4)})^{acjl\gamma\delta}_{bdik\dot \gamma\dot\delta}
(p_1,p_2,p_3,p_4)&=&-(G^4)^{cajl\gamma\delta}
_{bdki\dot\delta\dot\gamma}
(-p_4,p_2,p_3,-p_1)\nonumber\\
&=&-(G^{(4)})^{aclj\delta\gamma}_{dbik\dot\gamma\dot\delta}
(p_1,-p_3,-p_2,p_4)
\nonumber\\
&=&(G^{(4)})^{calj\delta\gamma}_{dbki\dot\delta\dot\gamma}
(-p_4,-p_3,-p_2,-p_1)
\ear
In the approximation (\ref{E.1}) the second functional derivatives read
\bear\label{E.5}
&&\left(\Gamma^{(2)}_k[\psi]\right)^{aj\gamma}_{bi\dot\gamma}
(q,q')=\frac{\delta^2\Gamma_k[\psi]}{\delta\psi^b_{j\gamma}
(q')\delta\bar\psi_a
^{i\dot\gamma}(q)}=\left(G^{(2)}\right)^{a\gamma}
_{b\dot\gamma}(q)\delta^j_i
(2\pi)^d\delta(q-q')\\
&&+\int\frac{d^dp}{(2\pi)^d}\frac{d^dp'}{(2\pi)^d}
\bar\psi_c^{k\dot\delta}(p)
\psi^d_{l\delta}(p')\left(G^{(4)}\right)^{acjl\gamma\delta}
_{bdik\dot\gamma\dot\delta}(-q,q',-p',p)(2\pi)^d\delta(q'-q+p'-p)
\nonumber
\ear
\bear\label{E.5a}
\frac{\delta^2\Gamma_k[\psi]}{\delta\psi^b_{j\gamma}
(q)\delta\psi^d_{l\delta}(q')}&=&\frac{1}{2}\int\frac{d^dp}
{(2\pi)^d}\bar\psi^{i\dot\gamma}_a(q+q'-p)\bar\psi^{k\dot\delta}
_c(p)\nonumber\\
&&(G^4)^{acjl\gamma\delta}_{bdik\dot\gamma\dot\delta}(p-q-q',
q,-q',p)\ear
\bear\label{E.5b}
\frac{\delta^2\Gamma_k[\psi]}{\delta\bar\psi^{i\dot\gamma}_a
(q)\delta\bar\psi^{k\dot\delta}_c(q')}&=&\frac{1}{2}\int\frac{d^dp}
{(2\pi)^d}\psi_{j\gamma}^b(p)\psi^d_{l\delta}
(-p-q-q')\nonumber\\
&&(G^{(4)})^{acjl\gamma\delta}_{bdik\dot\gamma\dot\delta}(-q,p,
p+q+q',q')\ear
The effective propagator (including the infrared cutoff
$R_k^{(\psi)}$) can
now be expanded correspondingly.
We will use here the ansatz\footnote{The matrix $\bar\gamma$
generalizes $\gamma^5$ to arbitrary $d$.}
for the two-point function
\be\label{E.6}
(G^{(2)})^a_b(q)=Z_\psi(c_a(q)\slq+m_a(q)\bar\gamma)\delta^a_b\ee
Noting that in the order
$(\bar\psi\psi)^2$ one has to take into account also
the off-diagonal $(\psi\psi)$ and $(\bar\psi\bar\psi)$
matrix elements in $\Gamma_k^{(2)}$ one finds in this order:
\bear\label{E.7}
&&\left[\left(\Gamma_k^{(2)}[\psi]+R^{(\psi)}_k\right)^{-1}\right]^
{aj\dot\gamma}_{bi\gamma}(q,q')=Z^{-1}_\psi h^{(a)\ \dot\gamma}_{\ \gamma}
(q)\delta^a_b\delta^j_i(2\pi)^d\delta(q-q')\nonumber\\
&&-Z^{-2}_\psi\int\frac{d^dp}{(2\pi)^d}\frac{d^dp'}{(2\pi)^d}
h^{(a)\dot\eta}_{\ \gamma}(q)(G^{(4)})^{acjl\eta\delta}_{bdik\dot
\eta\dot\delta}(-q,q',-p',p)\nonumber\\
&&h_\eta^{(b)\dot\gamma}(q')\bar\psi^{k\dot\delta}_c(p)\psi^d_{l\delta}
(p')(2\pi)^d\delta(q'-q+p'-p)\nonumber\\
&&+Z_\psi^{-3}\int\frac{d^dp_1}{(2\pi)^d}..\frac{d^dp_4}{(2\pi)^d}
\sum_eh^{(a)\dot\eta}_{\ \gamma}(q)
\Bigl\{(G^{(4)})^{acml\eta\delta}_{edik\dot\eta\dot\delta}
(-q,q-p_1-p_2,-p_2,-p_1)\nonumber\\
&&h^{(e)\dot\varepsilon}_\eta(q-p_1-p_2)(G^{(4)})^{efjq\epsilon\vartheta}
_{bgmp\dot\epsilon
\dot\vartheta}(-q'+p_3+p_4,q',p_3,p_4)\nonumber\\
&&+\frac{1}{4}\left(G^{(4)}\right)^{aeql\vartheta\delta}_{gdim\dot
\eta\dot\varepsilon}(-q,-p_3,-p_2,p_2-p_3-q)\nonumber\\
&&h^{(e)\dot\varepsilon}_\eta(p_2-p_3-q)\left(G^{(4)}\right)
^{cfjm\varepsilon \eta}_{bekp{\dot\delta}{\dot\vartheta}}
(p_1,q',q'+p_1-p_4,p_4)\Bigr\}
h^{(b)\dot\gamma}_\epsilon(q')\nonumber\\
&&\bar\psi^{k\dot\delta}_c(-p_1)\psi^d_{l\delta}
(p_2)\bar\psi_f^{p\dot\vartheta}(p_4)\psi^g_{q\vartheta}(-p_3)
(2\pi)^d\delta(q-q'-p_1-p_2+p_3+p_4)\ear
where the cutoff-propagator $h$ reads
$(r_k\equiv r^{(\psi)}_k$ in this appendix)
\be\label{E.8}
h_\gamma^{(a)\dot\gamma}(q)=\frac{((c_a(q)+r_k(q))\slq+m_a(q)
\bar\gamma)_\gamma
^{\ \dot\gamma}}{(c_a(q)+r_k(q))^2q^2+m^2_a(q)}\ee
This yields the following quadratic and quartic pieces in $\gamma_\psi$
\be\label{E.9}
\gamma_\psi^{(2)}=-Z^{-1}_\psi\int\frac{d^dp}{(2\pi)^d}
\frac{d^dq}{(2\pi)^d}\sum_a(G^{(4)})^{acil\eta\delta}_{adik\dot\eta\dot
\delta}(-q,q,-p,p)
H^{(a)\dot\eta}_\eta(q)\bar\psi^{k\dot\delta}_c(p)\psi^d_{l\delta}(p)\ee
and
\bear\label{E.10}
&&\gamma_\psi^{(4)}=Z_\psi^{-2}\int\frac{d^dp_1}{(2\pi)^d}...
\frac{d^dp_4}{(2\pi)^d}\frac{d^dq}{(2\pi)^d}\sum_{a,e}H_\epsilon^{(a)\dot
\eta}(q)\Bigl\{h^{(e)\dot\varepsilon}_\eta(q-p_1-p_2)\nonumber\\
&&(G^{(4)})^{acml\eta\delta}_{edik\dot\eta\dot\delta}
(-q,q-p_1-p_2,-p_2,-p_1)
(G^{(4)})^{efiq\epsilon\vartheta}_{agmp\dot\epsilon\dot\vartheta}
(-q+p_3+p_4,q,p_3,p_4)\nonumber\\
&&+\frac{1}{4}h^{(e)\dot\varepsilon}_\eta(p_2-p_3-q)\left(G^{(4)}\right)
^{aeql\vartheta\delta}_{gdim\dot\eta\dot\varepsilon}
(-q,-p_3,-p_2,p_2-p_3-q)\nonumber\\
&&\left(G^{(4)}\right)^{cfim\varepsilon\eta}_{aekp\dot\delta
\dot\vartheta}(p_1,q,q+p_1-p_4,p_4)\Bigr\}\nonumber\\
&&\bar\psi_c^{k\dot\delta}(-p_1)\psi^d_{l\delta}(p_2)
\bar\psi_f^{p\dot\vartheta}
(p_4)\psi^g_{q\vartheta}(-p_3)(2\pi)^d\delta(p_1+p_2-p_3-p_4)
\nonumber\\
\ear
where we use the shorthand $(\eta_\psi=-\frac{\partial}{\partial t}
\ln Z_\psi)$
\bear\label{E.11}
&&H^{(a)\dot\eta}_{\ \eta}(q)=h_\eta^{(a)\dot\gamma}
(q)\left(\frac{\partial}
{\partial t}r_k(q)-\eta_\psi r_k(q)\right)
\slq_{\dot\gamma}^{\ \gamma}h^{(a)\dot\eta}_{\ \gamma}(q)\nonumber\\
&&=\left(\frac{\partial}{\partial t}r_k(q)-\eta_\psi r_k(q)\right)
\left[(c_a(q)+r_k(q))^2q^2+m^2_a(q)\right]^{-2}.\nonumber\\
&&\left[((c_a(q)+r_k(q))^2q^2-m_a^2(q))\slq+2m_a(q)
(c_a(q)+r_k(q))q^2\bar\gamma
\right]\ear
To be more concrete, we consider the truncation
\bear\label{E.12}
&&\Gamma_{k,4}^{(\psi)}=- Z^2_\psi\int\frac{d^dp_1}
{(2\pi)^d}...\frac{d^dp_4}{(2\pi)^d}
(2\pi)^d\delta(p_1+p_2-p_3-p_4)\nonumber\\
&&\Bigl\{\lambda_\sigma(p_1,p_2,p_3,p_4){\cal M}_\sigma+\lambda_\rho
(p_1,p_2,p_3,p_4){\cal M}_\rho+\nonumber\\
&&\lambda_p(p_1,p_2,p_3,p_4){\cal M}_p+\lambda_n
(p_1,p_2,p_3,p_4){\cal N}\Bigr\}\ear
with $\lambda_\sigma(-p_4,-p_3,-p_2,-p_1)=\lambda_\sigma(p_1,p_2,p_3,p_4)$
and similar for  $\lambda_\rho,\lambda_p,\lambda_n$. This yields
\bear\label{E.13}
&&Z^{-2}_\psi(G^{(4)})^{acjl\gamma\delta}_{bdik\dot\gamma\dot\delta}
(p_1,p_2,p_3,p_4)=\nonumber\\
&&\lambda_\sigma(p_1,p_2,p_3,p_4)\delta^a_d\delta^c_b
\delta^j_i\delta^l_k\left(\delta^\gamma_{\dot\gamma}\delta^\delta_
{\dot\delta}
-(\bar\gamma)_{\dot\gamma}^{\ \gamma}(\bar\gamma)_{\dot\delta}^
{\ \delta}\right)\nonumber\\
&&
-\lambda_\sigma(p_1,-p_3,-p_2,p_4)\delta^a_b\delta^c_d
\delta^l_i\delta^j_k\left(\delta_{\dot\gamma}
^\delta\delta^\gamma_{\dot\delta}-(\bar\gamma)_{\dot\gamma}^
{\ \delta}(\bar\gamma)
^{\ \gamma}_{\dot\delta}\right)\nonumber\\
&&-\frac{1}{2}\lambda_\rho(p_1,p_2,p_3,p_4)\delta^a_d\delta^c_b\delta^j_i
\delta^l_k
\left((\gamma_\mu)_{\dot\gamma}^{\ \gamma}
(\gamma^\mu)_{\dot\delta}^{\ \delta}+(\gamma_\mu\bar\gamma)_{\dot\gamma}
^{\ \gamma}(\gamma^\mu\bar\gamma)_{\dot\delta}^{\ \delta}\right)
\nonumber\\
&&+\frac{1}{2}\lambda_\rho(p_1,-p_3,-p_2,p_4)\delta^a_b
\delta^c_d\delta^l_i
\delta^j_k
\left((\gamma_\mu)_{\dot\gamma}^{\ \delta}
(\gamma^\mu)_{\dot\delta}^{\ \gamma}+(\gamma_\mu\bar\gamma)_{\dot\gamma}
^{\ \delta}(\gamma^\mu\bar\gamma)_{\dot\delta}^{\ \gamma}\right)
\nonumber\\
&&-\frac{1}{N_c}\lambda_p(p_1,p_2,p_3,p_4)\delta^a_d\delta^c_b\delta^l_i
\delta^j_k(\gamma_\mu)_{\dot\gamma}^{\ \delta}(\gamma^\mu)_{\dot\delta}
^{\ \gamma}\nonumber\\
&&+\frac{1}{N_c}\lambda_p(p_1,-p_3,-p_2,p_4)\delta^a_b\delta^c_d\delta^j_i
\delta^l_k(\gamma_\mu)_{\dot\gamma}^{\ \gamma}(\gamma^\mu)_{\dot\delta}
^{\ \delta}\nonumber\\
&&+2\lambda_n(p_1,p_2,p_3,p_4)\delta^a_d\delta^c_b(T_z)_i^
{\ l}(T^z)_k^{\ j}
(\slp_1-\slp_3)_{\dot\gamma}^{\ \delta}(\slp_2-\slp_4)_{\dot\delta}^{\
\gamma}\nonumber\\
&&+2\lambda_n(p_1,-p_3,-p_2,p_4)\delta^a_b\delta^c_d(T_z)_i^
{\ j}(T^z)_k^{\ l}
(\slp_1+\slp_2)_{\dot\gamma}^{\ \gamma}(\slp_3+\slp_4)_
{\dot\delta}^{\
\delta}
\ear
{}From there we obtain
\bear\label{E.14}
&&\gamma^{(2)}_\psi=-\frac{1}{2}Z_\psi\int\frac{d^dp}{(2\pi)^d}
\frac{d^dq}{(2\pi)^d}\left(\frac{\partial}{\partial t}r_k(q)-\eta_\psi
r_k(q)\right)\cdot\nonumber\\
&&\sum_a[(c_a(q)+r_k(q))^2q^2+m^2_a(q)]^{-2}\nonumber\\
&&\Bigl[\bar\psi^i_a(p)\Bigl\lbrace-2^{\frac{d}{2}+2}N_c
m_a(q)(c_a(q)+r_k(q))q^2\bar\gamma\lambda_\sigma(-q,q,-p,p)\nonumber\\
&&-2^{\frac{d}{2}}N_c[(c_a(q)+r_k(q))^2q^2-m_a^2(q)]\slq\lambda_\rho
(-q,q,-p,p)
\nonumber\\
&&+\frac{2}{N_c}\Bigl[\left(2^{\frac{d}{2}}-2\right)\left((c_a(q)+r_k
(q))^2q^2-m^2_a(q)\right)\slq\\
&&+2^{\frac{d}{2}+1}m_a(q)(c_a(q)+r_k(q))q^2\bar\gamma\Bigr]\lambda_p
(-q,q-p,p)
\nonumber\\
&&+\frac{2(N_c^2-1)}{N_c}\quad \Bigl[((c_a(q)+r_k(q))^2q^2-m^2_a(q))
((p^2-q^2)\slq
+2q^2\slp-2(pq)\slp)\nonumber\\
&&+2m_a(q)(c_a(q)+r_k(q))q^2(q-p)^2\bar\gamma\Bigr]\lambda_n(-q,q,-p,p
)\Bigr\rbrace\psi^a_i(p)\nonumber\\
&&+[(c_a(q)+r_k(q))^2q^2-m^2_a(q)]\bar\psi^i_b(p)\slq\psi^b_i(p)
\nonumber\\
&&\left\lbrace-4\lambda_\sigma(-q,p,-q,p)+2(2-2^{\frac{d}{2}})\lambda_\rho
(-q,p,-q,p)+2^{\frac{d}{2}+1}\lambda_p(-q,p,-q,p)\right\rbrace\Bigr]
\nonumber\ear

For an investigation of $\gamma_\psi^{(4)}$ we first consider the
simple truncation (4.25)
which is relevant for sect. 4. This corresponds to
\bear\label{E.15}
&&\left(G^{(4)}\right)^{acjl\gamma\delta}_{bdik\dot\gamma\dot\delta}
(p_1,p_2,p_3,p_4)=2Z^2_\psi\nonumber\\
&&\Bigl\{\lambda_m(p_1,p_2,p_3,p_4)\delta^d_a\delta^b_c(T^z)_i^{\
l}(T_z)_k
^{\ j}(\gamma_\mu)_{\dot\gamma}^{\ \dot\delta}(\gamma^\mu)^{\ \gamma}
_{\dot\delta}\nonumber\\
&&-\lambda_m(p_1,-p_3,-p_2,p_4)\delta^b_a\delta^d_c(T^z)_i^{\ j}
(T_z)^{\ l}_k(\gamma_\mu)_{\dot\gamma}^{\ \gamma}(\gamma^\mu)^{\ \delta}
_{\dot\delta}\Bigr\}\ear
We also specialize to massless quarks and $c_a(q)=c(q)$.
Furthermore, we will omit from now on the second term in
the bracket in (\ref{E.10}). (The index structure of this
term is different and its contribution to $\gamma_\psi^{(4)}$
can be computed in a similar manner.)
One obtains
\bear\label{E.16}
&&\gamma_\psi^{(4)}=Z^2_\psi\int\frac{d^dp_1}{(2\pi)^d}
...\frac{d^dp_4}{(2\pi)^d}\frac{d^dq}{(2\pi)^d}(2\pi)^d\delta
(p_1+p_2-p_3-p_4)\nonumber\\
&&\Biggl[2^{\frac{d}{2}+1}N_F\lambda_m(-q,-p_3,-q+p_1-p_3,-p_1)\lambda_m
(-q-p_2+p_4,p_2,-q,p_4)\tilde H(q)\nonumber\\
&&\tilde h(q-p_1+p_3)(2q_\mu q_\nu-(p_1-p_3)_\mu q_\nu-(p_1-p_3)_\nu
q_\mu+((p_1q)-(p_3q)-q^2)\delta_{\mu\nu})\nonumber\\
&&\{\bar\psi(-p_1)\gamma^\mu T_z\psi(-p_3)\}\{\bar\psi(p_4)\gamma^\nu T^z
\psi(p_2)\}\nonumber\\
&&-4\lambda_m(-q,q-p_1-p_2,-p_2,-p_1)\lambda_m(-q+p_3+p_4,q,p_3,p_4)
\nonumber\\
&&\{\bar\psi(-p_1)\gamma^\mu\tilde h(q-p_1-p_2)(\slq-\slp_1-\slp_2)\gamma
_\nu T_zT^y\psi(-p_3)\}\nonumber\\
&&\{\bar\psi(p_4)\gamma^\nu\tilde H(q)\slq\gamma_\mu T_y T^z\psi(p_2)\}
\nonumber\\
&&-4\lambda_m(-q,q-p_1+p_3,p_3,-p_1)\lambda_m(-q-p_2+p_4,p_2,-q,p_4)
\nonumber\\
&&\{\bar\psi(-p_1)\gamma^\nu\tilde h(q-p_1+p_3)(\slq-\slp_1+\slp_3)
\gamma_\mu
\tilde H(q)\slq\gamma_\nu T_y T^zT^y\psi(-p_3)\}\nonumber\\
&&\{\bar\psi(p_4)\gamma^\mu T_z\psi(p_2)\}\nonumber\\
&&-4\lambda_m(-q,-p_3,-q+p_1-p_3,-p_1)\lambda_m(-q-p_2+p_4,q,-p_2,p_4)
\nonumber\\
&&\{\bar\psi(-p_1)\gamma^\mu T_z\psi(-p_3)\}\nonumber\\
&&\{\bar\psi(p_4)\gamma^\nu\tilde H(q)\slq\gamma_\mu\tilde h(q-p_1+p_3)
(\slq-\slp_1+\slp_3)\gamma_\nu T_y T^zT^y\psi(p_2)\}\Bigr]\ear
where $N_F$ is the number of quark flavours and
\bear\label{E.17}
&&H(q)=\tilde H(q)\slq,\quad h(q)=\tilde h(q)\slq\nonumber\\
&&\tilde h^{-1}(q)=(c(q)+r_k(q))q^2\nonumber\\
&&\tilde H(q)=q^2\tilde h^2(q)\left(\frac{\partial}{\partial t}r_k
(q)-\eta_\psi r_k(q)\right)=-\tilde\partial_t\tilde h(q)\ear
For the more general ansatz (E.12) the contribution to the flow
equation of the four-point function is rather lengthy and may be split
into
different terms
\bear\label{E.18}
&&\gamma_\psi^{(4)}=Z^2_\psi\int\frac{d^dp_1}{(2\pi)^d}...
\frac{d^dp_4}{(2\pi)^d}\frac{d^dq}{(2\pi)^d}
(2\pi)^d\delta(p_1+p_2-p_3-p_4)
\nonumber\\
&&\left\{A_{\sigma\sigma}+A_{\sigma\rho}+A_{\sigma p}+A_{\sigma n}+
A_{\rho\rho}+A_{\rho p}+A_{\rho n}+A_{pp}+A_{pn}+A_{nn}\right\}\ear
with $A_{\sigma\rho}\sim\lambda_\sigma\lambda_\rho$ etc. As an
example we give (again omitting the second term in the bracket in
(\ref{E.10}))
\bear\label{E.19}
&&A_{\sigma\sigma}=N_c\lambda_\sigma(-q,q-p_1-p_2,-p_2,-p_1)
\lambda_\sigma
(-q+p_3+p_4,q,p_3,p_4)\nonumber\\
&&\sum_{a,b}\Bigl[\{\bar\psi_a^i(-p_1)\psi^b_i(p_2)\}\{\bar\psi^j_b(p_4)
\psi^a_j(-p_3)\}tr(h^{(a)}(q-p_1-p_2)H^{(b)}(q))\nonumber\\
&&+\{\bar\psi^i_a(-p_1)\bar\gamma\psi^b_i(p_2)\}\{\bar\psi^j_b(p_4)
\bar\gamma
\psi^a_j(-p_3)\}tr(h^{(a)}(q-p_1-p_2)\bar\gamma H^{(b)}(q)\bar\gamma)
\nonumber\\
&&-\{\bar\psi^i_a(-p_1)\psi^b_i(p_2)\}\{\bar\psi^j_b(p_4)\bar\gamma
\psi^a_j(-p_3)\}tr(h^{(a)}(q-p_1-p_2)\bar\gamma H^{(b)}(q))\nonumber\\
&&-\{\bar\psi^i_a(-p_1)\bar\gamma\psi^b_i(p_2)\}\{\bar\psi^j_b(p_4)
\psi^a_j(-p_3)\}tr(h^{(a)}(q-p_1-p_2)H^{(b)}(q)\bar\gamma)\Bigr]
\nonumber\\
&&-\lambda_\sigma(-q,p_2,-q+p_1+p_2,-p_1)\lambda_\sigma
(-q+p_3+p_4,-p_3,-q,p_4)
\nonumber\\
&&\sum_c\Bigl[\left\{\bar\psi_a^i(-p_1)h^{(c)}(q-p_1-p_2)\psi_i^b(-p_3)
\right\}\left\{\bar\psi^j_b(p_4)H^{(c)}(q)\psi^a_j(p_2)\right\}
\nonumber\\
&&+\left\{\bar\psi_a^i(-p_1)\bar\gamma h^{(c)}(q-p_1-p_2)\bar\gamma
\psi_i^b(-p_3)\right\}\left\{\bar\psi_b^j(p_4)
\bar\gamma H^{(c)}(q)\bar\gamma \psi^a_j(p_2)\right\}\nonumber\\
&&-\left\{\bar\psi^i_a(-p_1)h^{(c)}(q-p_1-p_2)\bar\gamma\psi^b_i(-p_3)
\right\}\left\{\bar\psi^j_b(p_4)\bar\gamma H^{(c)}(q)\psi^a_j(p_2)\right\}
\nonumber\\
&&-\left\{\bar\psi^i_a(-p_1)\bar\gamma h^{(c)}(q-p_1-p_2)\psi^b_i(-p_3)
\right\}\left\{\bar\psi^j_b(p_4) H^{(c)}(q)\bar\gamma
\psi^a_j(p_2)\right\}
\Bigr]\nonumber\\
&&-\lambda_\sigma(-q,q-p_1-p_2,-p_2,-p_1)\lambda_\sigma(-q+p_3+p_4,-p_3,
-q,-p_4)\nonumber\\
&&\sum_a\Bigl[\left\{\bar\psi^i_a(-p_1)\psi^a_i(p_2)\right\}
\left\{\bar\psi^j_b(p_4)H^{(a)}(q)h^{(a)}(q-p_1-p_2)\psi^b_j(-p_3)\right\}
\nonumber\\
&&+\left\{\bar\psi^i_a(-p_1)\bar\gamma\psi^a_i(p_2)\right\}
\left\{\bar\psi^j_b
(p_4)\bar\gamma H^{(a)}(q)\bar\gamma h^{(a)}(q-p_1-p_2)\bar\gamma
\psi^b_j(-p_3)\right\}\nonumber\\
&&-\left\{\bar\psi^i_a(-p_1)\psi^a_i(p_2)\right\}\left\{\bar\psi^j_b(p_4)
\bar\gamma H^{(a)}(q)h^{(a)}(q-p_1-p_2)\bar\gamma\psi^b_j(-p_3)
\right\}\nonumber\\
&&-\left\{\bar\psi^i_a(-p_1)\bar\gamma\psi^a_i(p_2)\right\}\left\{\bar\psi
^j_b(p_4)H^{(a)}(q)\bar\gamma
h^{(a)}(q-p_1-p_2)\psi^b_j(-p_3)\right\}\Bigr]
\nonumber\\
&&-\lambda_\sigma(-q,p_2,-q+p_1+p_2,-p_1)\lambda_\sigma
(-q+p_3+p_4,q,p_3,p_4)
\nonumber\\
&&\sum_b\Bigl[\left\{\bar\psi^i_a(-p_1)h^{(b)}(q-p_1-p_2)H^{(b)}(q)
\psi^a_i(p_2)\right\}\left\{\bar\psi^j_b(p_4)\psi^b_j(-p_3)\right\}
\nonumber\\
&&+\left\{\bar\psi^i_a(-p_1)\bar\gamma h^{(b)}(q-p_1-p_2)\bar\gamma
H^{(b)}
(q)\bar\gamma\psi^a_i(p_2)\right\}\left\{\bar\psi^j_b(p_4)
\bar\gamma\psi_j^b
(-p_3)\right\}\nonumber\\
&&-\left\{\bar\psi^i_a(-p_1)h^{(b)}(q-p_1-p_2)\bar\gamma
H^{(b)}(q)\psi^a_i
(p_2)\right\}\left\{\bar\psi^j_b(p_4)\bar\gamma \psi_j^b(-p_3)\right\}
\nonumber\\
&&-\left\{\bar\psi^i_a(-p_1)\bar\gamma h^{(b)}(q-p_1-p_2)H^{(b)}(q)
\bar\gamma\psi^a_i(p_2)\right\}\left\{\bar\psi_b^j(p_4)
\psi^b_j(-p_3)\right\}\Bigr]
\ear
Here the trace tr is over Dirac spinor indices. Other terms like
$A_{\sigma\rho}$ and $A_{\rho\rho}$ can be obtained by inserting
$\gamma_\mu$ in appropriate places and accounting for appropriate
factors $1/2$ or $1/4$. In the chiral limit these expressions simplify
since for $m_a=0,\ c_a(q)=c(q)$ one has $h(q)=\tilde h(q){\slq}$ and
$H(q)=\tilde H(q){\slq}$. One finds
\bear\label{E.20}
&&A_{\sigma\sigma}=-2^{\frac{d}{2}+1}N_c\frac{q^2-(p_1 q)-(p_2
q)}{q^2(q-p_1-p_2)^2}\left(\frac{\partial}{\partial t} r_k(q)-
\eta_\psi r_k(q)\right)\nonumber\\
&&[c(q-p_1-p_2)+r_k(q-p_1-p_2)]^{-1}[c(q)+r_k(q)]^{-2}\nonumber\\
&&\lambda_\sigma(-q,q-p_1-p_2,-p_2,-p_1)
\lambda_\sigma(-q+p_3+p_4,q,p_3,p_4)
{\cal M}_\sigma(p_1,p_2,p_3,p_4)\nonumber\\
&&-2N_F\lambda_\sigma(-q,-p_3,-q+p_1-p_3,-p_1)\lambda_\sigma
(-q-p_2+p_4,p_2,-q,p_4)\nonumber\\
&&\Bigl[\left\{\bar\psi^i_a(-p_1)h(q-p_1+p_3)
\psi^b_i(p_2)\right\}\left\{
\bar\psi^j_b(p_4)H(q)\psi^a_j(-p_3)\right\}\nonumber\\
&&+\left\{\bar\psi^i_a(-p_1)h(q-p_1+p_3)\bar\gamma\psi^b_i(p_2)\right\}
\left\{\bar\psi^j_b(p_4)H(q)\bar\gamma\psi^a_j(-p_3)\right\}\Bigr]
\ear
where we have made for the second term a replacement $p_2\leftrightarrow
-p_3$ consistent with (\ref{E.19}). The first term gives a contribution
to the flow of $\lambda_\sigma$ whereas the second term, upon performing
the $q$-integration, contributes to   the  evolution of $\lambda_\rho$.
We observe that the omitted term in $A_{\sigma\sigma}$ arising from
the second term in the bracket in (\ref{E.10}) contributes only to the
flow of $\lambda_p$.

\section*{Appendix F: Evolution equation for the gauge field propagator}
\setcounter{equation}{0}
\renewcommand{\theequation}{F.\arabic{equation}}
In this appendix we compute the evolution equation for the gauge field
propagator in a pure Yang Mills theory without fermions. We use the gauge
fixing
\be\label{F.1}
S_{\rm gf}=\frac{1}{2\alpha}\int d^dx(\partial^\mu A^z_\mu)^2\ee
for arbitrary $\alpha$. We start with the gauge field-dependent
inverse propagator in momentum space
\bear\label{F.2}
&&(\tilde\Gamma_k^{(2)})^{y\mu}_{\nu z}(q',q)=\tilde Z_FP^\mu_\nu(q)
\delta^y_z(2\pi)^d
\delta(q'-q)\nonumber\\
&&+i\tilde g\tilde Z_Ff_z^{\ yw}\{(2q-q')_\nu A^\mu_w(q-q')\nonumber\\
&&+(2q'-q)^\mu A_{w\nu}(q-q')-(q+q')_\sigma A^\sigma_w(q-q')
\delta^\mu_\nu\}\nonumber\\
&&+\frac{1}{2}\tilde g^2\tilde Z_F\int\frac{d^dp}{(2\pi)^d}
\{(f_{zx}^{\ \ u}
f^y_{\ wu}+f_{zw}^{\ \ u}f^y_{\ xu}))\delta^\mu_\nu A^x_\sigma
(q-q'+p)A^{w\sigma}(-p)\nonumber\\
&&+2(f_z^{\ yu}f_{xwu}+f_{zw}^{\ \ u}f_{x\ u}^{\ y})A^{x\mu}(q-q'+p)
A^w_\nu(-p)
\}\ear
Here $P^\mu_\nu(q)$ is the general kinetic operator
\be\label{F.3}
P^\mu_\nu(q)=g(q)\delta^\mu_\nu+h(q)q_\nu q^\mu\ee
with $g$ and $h$ functions of $q^2$. In the classical approximation
one would
have
\be\label{F.4}
g(q)=q^2,\quad h(q)=(\frac{1}{\alpha}-1)\ee
Here we incorporate an infrared cutoff $\sim k$ such that for
massless gluons
\be\label{F.5}
\lim_{q^2\to0}g(q)\sim k^2,\quad \lim_{q^2\to0}h(q)\sim\frac{k^2}{q^2}.\ee
For the present calculation we keep $g$ and $h$ as free functions. Whereas
the $A$-dependent terms in (\ref{F.2}) can be derived from a term in the
action $\sim\frac{1}{4}\tilde Z_F F^z_{\mu\nu}F^{\mu\nu}_z$, we consider
here first the most general form of the $A$-independent term.

The evolution equation for the $k$-dependent effective action for gluons
$(\Gamma_k^{(A)}=\Gamma_k[A,\bar A=0])$
\be\label{F.6}
\frac{\partial}{\partial t}\Gamma_k^{(A)}=\frac{1}{2}\int\frac{d^dq}
{(2\pi)^d}\left\lbrace
\frac{\partial}{\partial t}[R_k(q)]^\nu_\mu[(\tilde\Gamma_k^{(2)}
)^{-1}]^{z\mu}_{\nu z}
(q,q)\right\rbrace-\epsilon=\gamma_A-\epsilon\ee
contains contributions from the gauge-field fluctuations $(\gamma_A)$
and the
ghosts $(\epsilon)$. They can be identified with corresponding
contributions
in the flow equation (4.18) for the quark-effective action in the
limit
of a truncation where $(\tilde\Gamma_k^{(2)})^{\alpha}_{\ \alpha'}$
does not
explicitly depend on $\psi$. Defining formally a partial derivative
$\tilde\partial_t$ acting only on $P^\mu_\nu$ with
\be\label{F.7}
\tilde\partial_t P^\mu_\nu(q)=\tilde Z_F^{-1}\frac{\partial}{\partial t}
\left(R_k(q)\delta^\mu_\nu+\tilde R_k(q) q_\nu q^\mu\right)
\ee
we can write $\gamma_A$ as  the derivative of a one-loop expression
\be\label{F.8}
\gamma_A=\frac{1}{2} Tr\tilde\partial_t\ln\tilde\Gamma_k^{(2)}\ee
The evolution of the gauge-field propagator can be extracted from
the term
quadratic in $A$ in $\gamma_A$
\be\label{F.9}
\gamma_A^{(2)}=\frac{1}{2}\int\frac{d^dp}{(2\pi)^d}\left\lbrace V_A(p)
A^\mu_z(-p)A_\mu^z(p)+W_A(p)p_\nu A^\nu_z(-p)p^\mu A^z_\mu(p)
\right\rbrace\ee
Our aim is the computation of the functions $V_A$ and $W_A$ which
depend on $p^2$. The ghost contributions will be added later.

We expand
\be\label{F.10}
Tr\tilde\partial_t\ln(\tilde P+\Delta)=Tr\tilde\partial_t\ln\tilde P
+Tr\tilde\partial_t(\Delta\tilde P^{-1})-\frac{1}{2} Tr\tilde\partial_t
(\Delta\tilde P^{-1}\Delta\tilde P^{-1})+...
\ee
where $\Delta$ is the $A$-dependent piece in $\tilde\Gamma_k^{(2)}$
(\ref{F.2}). The first term amounts to an irrelevant constant and the
contribution linear in $A$ from the second term vanishes due to
$f_z^{\ zw}=0$. We obtain more explicitly
\bear\label{F.11}
&&\gamma_A^{(2)}=\frac{1}{2}\int\frac{d^dq}{(2\pi)^d}\tilde\partial_
t\left\lbrace\Delta^{z\mu}_{\nu z}(q,q)\tilde Z_F^{-1}
(P^{-1})^\nu_\mu(q)\right\rbrace\\
&&-\frac{1}{4}\int\frac{d^dq}{(2\pi)^d}\frac{d^dp}{(2\pi)^d}
\tilde\partial
_t\left\lbrace\Delta^{y\mu}_{\nu z}(q,p)\tilde Z_F^{-1}(P^{-1})
^\rho_\mu (p)\Delta^{z\sigma}_{\rho y}(p,q)\tilde Z_F^{-1}(P^{-1})
^\nu_\sigma(q)\right\rbrace\nonumber
\ear
with
\bear\label{F.12}
(P^{-1})_\mu^\nu(q)&=&\frac{1}{g(p)}\left(\delta^\nu_\mu-\frac{h(q)}
{g(q)+h(q)q^2}q^\nu q_\mu\right)\nonumber\\
&=&\frac{1}{g(q)}\left(\delta^\nu_\mu-b(q)\frac{q^\nu q_\mu}{q^2}
\right)\ear
or
\be b(q)=\frac{h(q)q^2}{g(q)+h(q)q^2}\ee
Using the identity\footnote{The value $C=N_c$ obtains for the gauge
group $SU(N_c)$. We keep in this appendix the Casimir operator $C$ for
general nonabelian simple gauge groups.}
\be\label{F.14}
f_{zyw}f^{zyx}=(T_zT^z)^{\ x}_w=C\delta^x_w=N_c\delta^x_w\ee
this yields
\bear\label{F.15}
&&\gamma_A^{(2)}=\frac{1}{2}C\tilde g^2\int\frac{d^dp}{(2\pi)^d}
\frac{d^dq}{(2\pi)^d}\tilde\partial_t\left\lbrace g^{-1}(q)
\Bigl[(d-1-b(q))\right.A^z_\mu(p)A^\mu_z(-p)\nonumber\\
&&+\frac{b(q)}{q^2}q^\mu q_\nu A^z_\mu(p)A^\nu_z(-p)
\Bigr]\\
&&-\frac{1}{2}g^{-1}(q)g^{-1}(q+p)\left[(5p^2+2(pq)+2q^2)A^z_\mu(p)A^\mu_z
(-p)\right.\nonumber\\
&&+((4d-6)q^\mu q_\nu+(2d-3)(q^\mu p_\nu+p^\mu q_\nu)+(d-6)p^\mu p_\nu)
A^z_\mu(p)A^\nu_z(-p)\nonumber\\
&&-2\frac{b(q)}{q^2}\left\lbrace(q^2+2(pq))^2A^\mu_z(p)A^z_\mu(-p)
\right.\nonumber\\
&&+\left.[(p^2-2(pq)-q^2)q^\mu q_\nu-(q^2+3(pq))(q^\mu p_\nu+p^\mu q_\nu)
+q^2p^\mu p_\nu]A^z_\mu(p)A^\nu_z(-p)\right\rbrace\nonumber\\
&&+\frac{b(q)}{q^2}\frac{b(q+p)}{(q+p)^2}\left[(p^2)^2 q^\mu q_\nu-p^2(pq)
(q^\mu p_\nu+p^\mu q_\nu)
\left.\left.+(pq)^2p^\mu p_\nu\right] A^z_\mu(p)A^\nu_z(-p)\right]
\right\rbrace
\nonumber
 \ear
The term $\sim A^z_\mu(p)A^\nu _z(-p)$ can be projected on
contributions $\sim G_A$ using as a projector $\frac{1}{d-1}
\left(\delta^\nu_\mu-\frac{p_\mu p^\nu}{p^2}\right)$, and similarly
for contributions $\sim H_A$ with the projector $\frac{1}{d-1}
\frac{1}{p^2}\left( d\frac{p_\mu p^\nu}{p^2}-\delta^\nu_\mu\right)$.
With
\bear\label{F.16}
V_A(p)&=&C\tilde g^2\int\frac{d^dq}{(2\pi)^d}
\tilde\partial_tv_A(p,q)\nonumber\\
W_A(p)&=&C\tilde g^2\int\frac{d^dq}{(2\pi)^d}\tilde\partial_tw_A(p,q)\ear
we obtain
\bear\label{F.17}
&&v_A(p,q)=g^{-1}(q)\left[d-1-b(q)+\frac{b(q)}{d-1}\left(1-\frac{(pq)^2}
{q^2p^2}\right)\right]\nonumber\\
&&-\frac{1}{2}g^{-1}(q)g^{-1}(q+p)\left[5p^2+2(pq)+2q^2+\frac{4d-6}{d-1}
\left(q^2-\frac{(pq)^2}{p^2}\right)\right.\nonumber\\
&&-\frac{2}{d-1}\frac{b(q)}{q^2}\left\lbrace(d-2)(q^2)^2+(4d-6)q^2(pq)
+q^2p^2+(4d-5)(pq)^2\right.\nonumber\\
&&+2\frac{(pq)^3}{p^2}+\frac{(pq)^2q^2}{p^2}\Bigr\rbrace
\left.
+\frac{1}{d-1}\frac{b(q)}{q^2}\frac{b(q+p)}{(q+p)^2}
\left\lbrace(p^2)^2q^2-(pq)^2p^2
\right\rbrace\right]\ear
and
\bear\label{F.18}
&&w_A(p,q)=g^{-1}(q)\frac{b(q)}{q^2}\left[\frac{d}{d-1}\frac{(pq)^2}{(p^2)
^2}-\frac{1}{d-1}\frac{q^2}{p^2}\right]\nonumber\\
&&-\frac{1}{2}g^{-1}(q)g^{-1}(q+p)\left[\frac{4d-6}{d-1}
\left(d\frac{(pq)^2}
{(p^2)^2}-\frac{q^2}{p^2}\right)+(4d-6)\frac{(pq)}{p^2}+d-6
\right.\nonumber\\
&&-\frac{2b(q)}{q^2}\left\lbrace\frac{d-2}{d-1}q^2-2\frac{d-2}{d-1}
\frac{
q^2(pq)}{p^2}\right.\nonumber\\
&&\left.-\frac{5d-6}{d-1}\frac{(pq)^2}{p^2}+\frac{1}{d-1}
\frac{(q^2)^2}{p^2}-
\frac{d}{d-1}\frac{(pq)^2q^2}{(p^2)^2}-\frac{2d}{d-1}\frac{(pq)^3}
{(p^2)^2}\right\rbrace\nonumber\\
&&\left.+\frac{1}{d-1}\frac{b(q)}{q^2}\frac{b(q+p)}{(q+p)^2}\lbrace(pq)^2-
q^2p^2\rbrace\right]\ear

We next evaluate the contribution from the ghosts given by $\epsilon$
(\ref{3.17}). With the approximation (\ref{3.18}) and $\bar A=0$ one has
\be\label{F.19}
\epsilon=Tr\left\lbrace\left(\frac{\partial}{\partial t}R_k^{({\rm gh})}
\right)
\left(-\partial^\mu D_\mu[A]+R_k^{({\rm gh})}\right)^{-1}\right\rbrace\ee
where $R^{({\rm gh})}$ is now a function of the normal Laplacian.
We generalize
this formula by taking into account the most general effective ghost
propagator.
This corresponds in momentum space to the truncation
\be\label{F.20}
\left[\Gamma_k^{({\rm gh})(2)}+R_k^{({\rm gh})}\right]^y_{\ z}(q',q)=
Z_{\rm gh}
P_{\rm gh}(q)\delta^y_z(2\pi)^d\delta(q-q')-i\tilde g Z_{\rm gh}
f_{w\ z}^{\ y}
A^w_\mu(q'-q){q'}^\mu\ee
An expansion of the propagator up to quadratic order in $A$ gives
\bear\label{F.21}
&&\left[\left(\Gamma_k^{({\rm gh})(2)}+R_k^{({\rm gh})}\right)^{-1}
\right]^y_
{\ z}(q',q)=Z^{-1}_{\rm gh}P^{-1}_{\rm gh}(q)\delta^y_z(2\pi)^d
\delta(q-q')
\nonumber\\
&&+i\tilde gZ^{-1}_{\rm gh}P^{-1}_{\rm gh}(q')P^{-1}_{\rm gh}(q)
f_{w\ z}^{\ y}
{q'}^\mu A^w_\mu(q'-q)\nonumber\\
&&-\tilde g^2Z^{-1}_{\rm gh}P^{-1}_{\rm gh}(q')P^{-1}_{\rm gh}(q)
f_{w\ x}^{\ y}
f^{\ x}_{v\ z}\cdot\nonumber\\
&&\int\frac{d^dp}{(2\pi)^d}P^{-1}_{\rm gh}(p){q'}^\mu p^\nu A^w_\mu
(q'-p)A^v_\nu(p-q)\ear
This yields, up to an irrelevant constant
\bear\label{F.22}
\epsilon&=&C\tilde g^2\int\frac{d^dq}{(2\pi)^d}Z^{-1}_{\rm gh}
\frac{\partial}{\partial t}R_k^{({\rm gh})}(q)P^{-2}_{\rm gh}(q)\cdot
\nonumber\\
&&\int\frac{d^dp}{(2\pi)^d}P^{-1}_{\rm gh}(q-p)q^\mu(q-p)_\nu A^z_\mu(p)
A^\nu_z(-p)\ear
With
\be\label{F.23}
\epsilon=-\frac{1}{2}\int\frac{d^dp}{(2\pi)^d}\left\lbrace
V_{\rm gh}(p)\delta^\mu_\nu+W_{\rm gh}(p)p^\mu p_\nu\right\rbrace A^z_\mu
(p)A^\nu_z(-p)\ee
and defining $g_{\rm gh}$ and $h_{\rm gh}$ in analogy to (\ref{F.16})
one finds
\be\label{F.24}
V_{\rm gh}=C\tilde g^2\frac{1}{d-1}\left(\delta_{\mu\nu}-
\frac{p_\mu p_\nu}{p^2}\right)\int\frac{d^dq}{(2\pi)^d}
\tilde\partial_ t\left\lbrace P^{-1}_{\rm gh}(q)P^{-1}_{\rm gh}
(q+p)q^\mu(q+p)^\nu\right\rbrace
\ee
\be\label{F.25}
v_{\rm gh}=\frac{1}{d-1}P^{-1}_{\rm gh}(q)P^{-1}_{\rm gh}(q+p)
\left(q^2-\frac{(pq)^2}{p^2}\right)\ee
\be\label{F.26}
W_{\rm gh}=C\tilde g^2\frac{1}{d-1}\frac{1}{p^2}\left(d\frac{p_\mu p_\nu}
{p^2}-\delta_{\mu\nu}\right)\int\frac{d^dq}{(2\pi)^d}\tilde\partial
_t\left\lbrace P^{-1}_{\rm gh}(q)P^{-1}_{\rm gh}(q+p)q^\mu
(q+p)^\nu\right\rbrace\ee
\be\label{F.27}
w_{\rm gh}=\frac{1}{d-1}P^{-1}_{\rm gh}(q)P^{-1}_{\rm gh}(q+p)\left[
d\frac{(pq)^2}{(p^2)^2}+(d-1)\frac{pq}{p^2}-\frac{q^2}{p^2}\right]\ee

Combining the contributions from $\gamma_A$ and $\epsilon$ we arrive
at the flow equation for the gluon propagator for the pure Yang-Mills
theory.
It describes the $k$-dependence of the function $G_A(q)$ and $H_A(q)$
as defined in (\ref{5.1}) and reads
\bear\label{F.28}
&&\frac{\partial}{\partial t}G_A(p)=\frac{\partial}{\partial t}(\tilde Z
_Fg(p)-R_k(p))=V_A(p)+V_{\rm gh}(p)\nonumber\\
&&\frac{\partial}{\partial t}H_A(p)=\frac{\partial}{\partial t}(\tilde Z
_Fh(p)-\tilde R_k(p))=W_A(p)+W_{\rm gh}(p)\ear
We observe that $V_A$ and $W_A$ are defined in terms of $G_A$ and $H_A$
through the relation $\Gamma_k^{(2)}(0)=\tilde\Gamma_k^{(2)}(0)-R_k^{(A)},
\ G_A=\tilde Z_F g-R_k,\ H_A=\tilde Z_Fh-\tilde R_k$.
The system (\ref{F.28}) constitutes therefore a coupled system of partial
differential equations for the functions $G_A$ and $H_A$ which depend on
two variables, namely $k$ and $p^2$. It describes the scale-dependence
of the gluon propagator under the influence of the three- and four-gluon
interactions contained in $(F_{\mu\nu})^2$ and the corresponding ghost
contributions.

It is instructive to study first
a simplified version of this system where the
propagator $\tilde\Gamma_k^{(2)}$ is approximated on the r.h.s. by
\bear\label{F.29}
\tilde Z_Fg(q)&=&\tilde Z_Fq^2+R_k(q)=\tilde Z_F P(q),\
h(q)=0\\
&&P_{\rm gh}(q)=P(q),\ \tilde\partial_t P(q)=
\frac{\partial}{\partial t}P(q)
\ear
This corresponds to the classical approximation (\ref{F.4})
wit $\alpha=1$. It implies $b(q)=0$ and
\bear\label{F.30}
v_A(p,q)&=&-\frac{1}{2}P^{-1}(q)P^{-1}(q+p)
\left[\frac{6d-8}{d-1}q^2+2(pq)
+5p^2-\frac{4d-6}{d-1}\frac{(pq)^2}{p^2}\right]\nonumber\\
&&+(d-1)P^{-1}(q)\ear
\bear\label{F.31}
&&w_A(p,q)=-\frac{1}{2}P^{-1}(q)P^{-1}(q+p)\nonumber\\
&&\left[(4d-6)\frac{(pq)}{p^2}
+d-6+\frac{2d(2d-3)}{d-1}\frac{(pq)^2}{(p^2)^2}-\frac{4d-6}{d-1}
\frac{q^2}{p^2}\right]\ear
Combining this with the ghost contribution one obtains
\bear\label{F.32}
&&\frac{\partial}{\partial t}G_A(p)=(d-1)C\tilde g^2\int
\frac{d^dq}{(2\pi)^d}
\frac{\partial}{\partial t}P^{-1}(q)\\
&&-\frac{1}{2}C\tilde g^2 \int\frac{d^dq}{(2\pi)^d}
\frac{\partial}{\partial t}\left(P^{-1}(q)P^{-1}(q+p)\right)
\left[\frac{6d-10}{d-1}q^2+2(pq)+5p^2-4\frac{d-2}{d-1}
\frac{(pq)^2}{p^2}
\right]\nonumber\ear
\bear\label{F.33}
\frac{\partial}{\partial t}H_A(p)&=&
-\frac{1}{2}C\tilde g^2\int\frac{d^dq}{(2\pi)^d}
\frac{\partial}{\partial t}\left(P^{-1}(q)P^{-1}(q+p)\right)
\cdot\nonumber\\
&&\left[d-6+4(d-2)\frac{pq}{p^2}-4\frac{d-2}{d-1}\frac{q^2}{p^2}
+4d\frac{d-2}{d-1}\frac{(pq)^2}{(p^2)^2}
\right]\ear
If we want to study the behaviour of the gluon propagator for small values
of the momentum $p^2\ll k^2$ we can expand the functions $G_A$ and $H_A$
in powers of $p^2$. For this purpose we expand $P^{-1}(q+p)$ (with $\dot
P(q)
\equiv\frac{\partial}{\partial q^2}P(q))$
\bear\label{F.34}
&&P^{-1}(q+p)=P^{-1}(q)-2(pq)\dot P(q)P^{-2}(q)\nonumber\\
&&-p^2\dot P(q)P^{-2}(q)-2(pq)^2(\ddot P(q)P^{-2}(q)-2\dot P^2(q)P^{-3}
(q))+0(p^3)\ear
and use the identities
\bear\label{F.35}
&&\int\frac{d^dq}{(2\pi)^d}f(q^2)q_\mu
q_\nu=\frac{1}{d}\delta_{\mu\nu}\int
\frac{d^dq}{(2\pi)^d}q^2f(q^2)\\
&&\int\frac{d^dq}{(2\pi)^d}f(q^2)q_\mu q_\nu q_\rho
q_\sigma=\frac{1}{d(d+2)}(\delta_{\mu\nu}
\delta_{\rho\sigma}+\delta_{\mu\rho}
\delta_{\nu\sigma}+\delta_{\mu\sigma}\delta_{\nu\rho})
\int\frac{d^dq}{(2\pi)^d}q^4f(q^2)\nonumber\ear
This yields
\be\label{F.36}
\frac{\partial}{\partial t}G_A(p)=V^{(0)}+V^{(1)}p^2+O(p^4)\ee
with
\be\label{F.37}
V^{(0)}=C\tilde g^2\int\frac{d^dq}{(2\pi)^d}\frac{\partial}{\partial t}
\left\lbrace (d-1)P^{-1}(q)-\frac{3d-4}{d}q^2P^{-2}(q)\right\rbrace\ee
\bear\label{F.38}
&&V^{(1)}=-\frac{1}{2}C\tilde g^2\int\frac{d^dq}{(2\pi)^d}
\frac{\partial}{\partial t}
\left\lbrace 5P^{-2}(q)-
\frac{6d-4}{d}q^2\dot P(q)P^{-3}(q)\right.\nonumber\\
&&\left.-\frac{4(3d-2)}{d(d+2)}q^4[\ddot P(q)P^{-3}(q)-2
\dot P^2(q)P^{-4}(q)]\right\rbrace
\ear
and
\bear\label{F.39}
&&\frac{\partial}{\partial t}H_A(0)=W^{(0)}=-\frac{1}{2}C\tilde
g^2\int\frac{d^dq}{(2\pi)^d}\frac{\partial}{\partial t}\Biggl\lbrace
(d-6)P^{-2}(q)-\frac{8}{d}(d-2)q^2\dot P(q)P^{-3}(q)\nonumber\\
&&-\frac{16(d-2)}{d(d+2)}q^4[\ddot P(q)P^{-3}(q)-2\dot P^2(q)P^{-4}(q)]
\Biggr\rbrace\ear

With $x=q^2$ we define the constants $l_n^d,m^d_n,v_d$ by
\be\label{F.40}
l^d_n=-\frac{1}{2}k^{2n-d}\int^\infty_0dx\
x^{\frac{d}{2}-1}\frac{\partial}
{\partial t}P^{-n}\ee
\be\label{F.41}
m^d_n=-\frac{1}{2}k^{2n-d-2}\int^\infty_0dx\
x^{\frac{d}{2}}\frac{\partial}
{\partial t}\left\lbrace \dot P^2P^{-n}\right\rbrace\ee
\be\label{F.42}
v_d^{-1}=2^{d+1}\pi^{\frac{d}{2}}\Gamma(\frac{d}{2})\ee
and find
\be\label{F.43}
V^{(0)}=-4v_dC\tilde g^2k^{d-2}\left((d-1)l^d_1-\frac{3d-4}{d}
l_2^{d+2}\right)\ee
\bear\label{F.44}
V^{(1)}&=&v_dC\tilde g^2k^{4-d}\left(10l^d_2-\frac{8(3d-2)}
{d(d+2)}m^{d+2}_4
\right)\nonumber\\
&&+\frac{4(3d-2)}{d(d+2)}v_dC\tilde g^2\int dx\frac{d}{dx}
\frac{\partial}{\partial t}\left\lbrace
x^{\frac{d}{2}+1}\dot PP^{-3}\right\rbrace\ear
\bear\label{F.45}
W^{(0)}&=&2v_dC\tilde g^2k^{d-4}\left((d-6)l^d_2-
\frac{16(d-2)}{d(d+2)}
m^{d+2}_4\right)\nonumber\\
&&+16v_d\frac{(d-2)}{d(d+2)}C\tilde g^2\int dx \frac{d}{dx}
\frac{\partial}{\partial t}
\left\lbrace x^{\frac{d}{2}+1}\dot PP^{-3}\right\rbrace\ear
Using the definition (\ref{4.7}) for $P(q)$ one easily
verifies the identities
$l^{2n}_n=1,\ m^{2n-2}_n=1$.
For $d=2$ the coefficients $V^{(0)}$ and $W^{(0)}$ vanish and $V^{(1)}$
is proportional to $5l^2_2-4m^4_4=10\ln 2-2$. For $d=4$ one
has $v_4=1/32\pi^2$
and
\bear\label{F.46}
V^{(0)}&=&-\frac{C}{8\pi^2}\tilde g^2k^2(3l_1^4-2l^6_2)=0\nonumber\\
V^{(1)}&=&\frac{10}{3}C\frac{\tilde g^2}{16\pi^2}\nonumber\\
W^{(0)}&=&-\frac{10}{3}C\frac{\tilde g^2}{16\pi^2}\ear

A similar expansion for small values of $p^2$ can be made without the
approximations (\ref{F.29}). One replaces $P$ in (\ref{F.40}),
(\ref{F.41})
by $P_A$ or $P_{gh}$ and adds a contribution $\sim\partial\ln
\tilde Z_F/\partial t$ for the $t$-derivative of the factor
$\tilde Z_F$ in $R_k$.

For a study of the behaviour of (\ref{F.32}), (\ref{F.33}) for general
$p$ it is convenient to introduce in the definition of $P$ an explicit
ultraviolet cutoff $\Lambda$\footnote{The previous form of $P(q)$
obtains for $\Lambda\to\infty$.}
\be\label{F.47}
P^{-1}_{k,\Lambda}(q)=\frac{\exp\left(-\frac{q^2}{\Lambda^2}\right)-
\exp\left(-\frac{q^2}{k^2}\right)}{q^2}=\int^{k^{-2}}_{\Lambda^{-2}}
d\alpha\exp(-\alpha q^2)\ee
This permits to perform explicitly the momentum integration
\bear\label{F.48}
&&\frac{\partial}{\partial t}G_A(p)=(d-1)C\tilde g^2\frac{\partial}
{\partial t}
\int^{k^{-2}}_{\Lambda^{-2}}d\alpha\int\frac{d^dq}{(2\pi)^d}
\exp(-\alpha q^2)
\nonumber\\
&&-\frac{1}{2}C\tilde g^2\frac{\partial}{\partial t}
\int^{k^{-2}}_{\Lambda^{-2}}
d\alpha\int^{k^{-2}}_{\Lambda^{-2}}
d\beta\int\frac{d^dq}{(2\pi)^d}\exp(-\alpha q^2-\beta(q+p)^2)
\nonumber\\
&&\left[\frac{6d-10}{d-1}q^2+2(pq)+5p^2-4\frac{d-2}{d-1}
\frac{(pq)(pq)}{p^2}
\right]\nonumber\\
&&=-2(d-1)(4\pi)^{-\frac{d}{2}}C\tilde g^2k^{d-2}\nonumber\\
&&-\frac{1}{2}(4\pi)^{-\frac{d}{2}}C\tilde g^2\frac{\partial}
{\partial t}
\int^{k^{-2}}_{\Lambda^{-2}}d\alpha d\beta(\alpha+\beta)^{-\frac{d}{2}}.
\nonumber\\
&&\left(\frac{3d-4}{\alpha+\beta}+p^2\left(5-\frac{2\alpha\beta}
{(\alpha+\beta)^2}\right)\right)
\exp\left(-\frac{\alpha\beta}{\alpha+\beta}p^2\right)\ear
\bear\label{F.49}
&&\frac{\partial}{\partial t}H_A(p)=-\frac{1}{2}(4\pi)^{-\frac{d}{2}}
C\tilde g^2\frac{\partial}{\partial t}\int^{k^{-2}}_{\Lambda^{-2}}
d\alpha d\beta(\alpha+\beta)^{-\frac{d}{2}}
\nonumber\\
&&\left(d-6-4(d-2)\frac{\alpha\beta}{(\alpha+\beta)^2}\right)
\exp\left(
-\frac{\alpha\beta}{\alpha+\beta}p^2\right)\ear
Using the integral
\be\label{F.50}
\lim_{\Lambda\to\infty}\frac{d}{dt}\int^{k^{-2}}_{\Lambda^{-2}}d\alpha
\int^{k^{-2}}_{\Lambda^{-2}}d\beta(\alpha+\beta)^{-(y+2)}\exp\left(
-\frac{\alpha\beta}{\alpha+\beta}p^2\right)=-4k^{2y}\exp
\left(-\frac{p^2}{k^2}\right)\hat J(y)\ee
with
 \bear\label{F.51}
&&\hat J(y)=\int^1_{\frac{1}{2}}d\gamma \gamma^y\exp
\left(\frac{p^2}{k^2}\gamma
\right)\nonumber\\
&&\hat J(y+1)=k^2\frac{\partial}{\partial p^2}\hat J(y)\ear
and
\bear\label{52}
\hat J(0)&=&\frac{k^2}{p^2}\left(\exp\frac{p^2}{k^2}-\exp
\frac{p^2}{2k^2}\right)\nonumber\\
\hat J(1)&=&-\frac{k^4}{(p^2)^2}\left(\exp\frac{p^2}{k^2}
-\exp\frac{p^2}{2k^2}\right)
+\frac{k^2}{p^2}\left(\exp\frac{p^2}{k^2}-\frac{1}{2}\exp
\frac{p^2}{2k^2}
\right)\nonumber\\
\hat J(2)&=&2\frac{k^6}{(p^2)^3}\left(\exp\frac{p^2}{k^2}
-\exp\frac{p^2}{2k^2}\right)
-\frac{2k^4}{(p^2)^2}\left(\exp\frac{p^2}{k^2}-\frac{1}{2}
\exp\frac{p^2}{2k^2}
\right)\nonumber\\
&&+\frac{k^2}{p^2}\left(\exp\frac{p^2}{k^2}
-\frac{1}{4}\exp\frac{p^2}{2k^2}\right)\ear
we find
\bear\label{F.53}
\frac{\partial}{\partial t}G_A(p)&=&C\tilde g^2{\cal G}(p)\nonumber\\
{\cal G}(p)&=&-2(d-1)(4\pi)^{-\frac {d}{2}}k^{d-2}\nonumber\\
&&+2(4\pi)^{-\frac{d}{2}} k^{d-2}\exp\left(-\frac{p^2}{k^2}\right)
\left\lbrace(3d-4)\hat J(\frac{d}{2}-1)\right.\nonumber\\
&&\left.+\frac{p^2}{k^2}\left(5\hat J\left(\frac{d}{2}-2\right)-2
\hat J\left(\frac{d}{2}-1\right)+2\hat J\left(\frac{d}{2}\right)
\right)\right
\rbrace\ear
\bear\label{F.54}
\frac{\partial}{\partial t}H_A(p)&=&2(4\pi)^{-\frac {d}{2}}C\tilde
g^2k^{d-4}\exp\left(-\frac{p^2}{k^2}\right)
\left\lbrace(d-6)\hat J\left(\frac{d}{2}-2\right)\nonumber\right.\\
&&\left.-4(d-2)\hat J\left(\frac{d}{2}-1\right)+4(d-2)\hat J
\left(\frac{d}{2}\right)\right\rbrace\ear
In particular, this yields for $d=4$
\bear\label{F.55}
\frac{\partial}{\partial t}G_A(p)&=&\beta_A(p)\\
&=&\frac{C}{16\pi^2}\tilde g^2k^2\left\lbrace4+12\frac{k^2}{p^2}-8
\frac{k^4}{(p^2)^2}-\exp\left(-\frac{p^2}{2k^2}\right)
\left(9+8\frac{k^2}{p^2}-8\frac{k^4}{(p^2)^2}\right)\right\rbrace
\nonumber\ear
The qualitative behaviour of $\beta_A(p)$ can easily be understood
by expanding for $p^2\ll 2k^2$
\be\label{F.56}
\beta_A(p)=\frac{C}{16\pi^2}\tilde g^2p^2\left(\frac{10}{3}-
\frac{15}{16}\frac{p^2}{k^2}+\frac{79}{480}\left(
\frac{p^2}{k^2}\right)^2-\frac{61}{2880}\left(\frac{p^2}{k^2}\right)^3+
\frac{29}{13440}\left(\frac{p^2}{k^2}\right)^4-...\right)\ee
whereas for $p^2\gg 2k^2$ one has
\be\label{F.57}
\beta_A(p)=\frac{C}{16\pi^2}\tilde g^2k^2\left(4+12\frac{k^2}
{p^2}-8\left(\frac{
k^2}{p^2}\right)^2\right)\ee
It is instructive to write the gluon propagator in terms of
momentum-dependent wave function renormalizations $Z_t,Z_l$ for the
transversal and longitudinal gluons and a gluon mass term $\bar m^2_A$:
\be\label{F.58}
G_A(p)\delta_{\mu\nu}+H_A(p)p_\mu p_\nu=\bar m^2_A\delta_{\mu\nu}
+Z_t(p)(p^2\delta_{\mu\nu}-p_\mu p_\nu)+Z_l(p)p_\mu p_\nu\ee
\bear\label{F.59}
&&\bar m^2_A=G_A(0)\nonumber\\
&&Z_t(p)=\frac{G_A(p)-G_A(0)}{p^2}\nonumber\\
&&Z_l(p)=H_A(p)+\frac{G_A(p)-G_A(0)}{p^2}\ear
We may now define a $k$-dependent gauge-fixing parameter $\alpha(k)$
and $\tilde Z_F$ by
\bear\label{F.60}
&&\frac{1}{\alpha(k)}=Z_l(0)\nonumber\\
&&\tilde Z_F(k)=Z_t(0)\ear
where both definitions make only sense for $Z_l(0)$ and $Z_t(0)$ positive.
{}From (\ref{F.46}) we obtain
\bear\label{F.61}
&&\frac{\partial}{\partial t}\alpha=-\frac{1}{\alpha^2}
\left(W^{(0)}+V^{(1)}\right)=0\nonumber\\
&&\frac{\partial}{\partial t}\tilde Z_F=V^{(1)}=\frac{10}{3}
C\frac{g^2_k}{16\pi^2}\tilde Z_F\nonumber\\
&&\frac{\partial}{\partial t}\bar m^2_A=V^{(0)}=0\ear
where we should mention that the vanishing of $V^{(0)}$ is
particular to our choice of the cutoff function $R_k$. Even though
$\alpha(k)$ does not depend on $k$ in our approximation, the relevant
quantity for the validity of the approximation (\ref{F.29}) is the
ratio $Z_l(0)/Z_t(0)$. The renormalized gauge fixing parameter
\be\label{F.62}
\alpha_R(k)=\tilde Z_F(k)\alpha(k)\ee
obeys
\be\label{F.63}
\frac{\partial}{\partial t}\alpha_R=\frac{10}{3}C\frac{g^2_k}{16\pi^2}
\alpha_R\ee
and moves away from $\alpha_R=1$ appropriate for (\ref{F.29}). The
approximation (\ref{F.29}) will therefore break down for $k$ in
the vicinity of the confinement scale where $g^2_k$ becomes large.
We finally give the running of $\tilde g^2$ in terms of the $\beta$
function for the renormalized gauge coupling $g^2_k$
\be\label{F.64}
\frac{\partial}{\partial t}\tilde g^2=\tilde Z_F
\left(\beta_{g^2}+\frac{10}
{3}C\frac{g^4_k}{16\pi^2}\right)=-\frac{C}{4\pi^2}g^4_k\tilde Z_F\ee

For comparison, we have also computed the flow of the gluon
propagator for $\alpha=0$. We start by summarizing
the preceding results for arbitrary $\alpha$.
\bear\label{F.66}
&&\frac{\partial}{\partial t}G_A(q)=N_cg^2_k\tilde Z_F\int\frac{d^4q'}
{(2\pi)^4}\tilde\partial_t\Biggl\lbrace
\left(G_A(q')+R_k(q')\right)^{-1}\tilde Z_F\left[3-\frac{3}{4}
b(q')\right]\nonumber\\
&&-\frac{1}{2}\left(G_A(q')+R_k(q')\right)^{-1}
\left(G_A(q+q')+R_k(q+q')\right)^{-1}\tilde Z^2_F\nonumber\\
&&\left[5q^2+2(qq')+\frac{16}{3}{q'}^2-\frac{10}{3}\frac{(qq')^2}{q^2}
\right.\nonumber\\
&&-\frac{2}{3}b(q')\left(2{q'}^2+10(qq')+q^2+11\frac{(qq')^2}{{q'}^2}+2
\frac{(qq')^3}{q^2{q'}^2}+\frac{(qq')^2}{q^2}\right)\nonumber\\
&&\left.+\frac{1}{3}b(q')b(q+q')\frac{q^2}{(q+q')^2}\left(q^2-
\frac{(qq')^2}{{q'}^2}\right)\right]
\nonumber\\
&&+\frac{1}{3}P^{-1}_{\rm gh}(q')P^{-1}_{\rm gh}(q+q')
\left[{q'}^2-\frac{(qq')^2}{q^2}\right]\Biggr\rbrace\ear
and
\bear\label{F.67}
&&\frac{\partial}{\partial t}H_A(q)=-\frac{1}{2}N_cg^2_k\tilde
Z_F\int\frac{d^4q'}{(2\pi)^4}\tilde\partial_t\nonumber\\
&&\Biggl\lbrace\left(G_A(q')+R_k(q')\right)^{-1} \left(G_A(q+q')+
R_k(q+q')\right)^{-1}\tilde Z_F^2\nonumber\\
&&\left[\frac{40}{3}\frac{(qq')^2}{(q^2)^2}-
\frac{10}{3}\frac{{q'}^2}{q^2}+10\frac{(qq')}{q^2}-2
\right.\nonumber\\
&&-2b(q')\left(\frac{2}{3}-\frac{4}{3}\frac{(qq')}{q^2}-\frac{14}{3}
\frac{(qq')^2}{q^2{q'}^2}+\frac{1}{3}\frac{{q'}^2}{q^2}-\frac{4}{3}
\frac{(qq')^2}{(q^2)^2}-\frac{8}{3}\frac{(qq')^3}{(q^2)^2{q'}^2}
\right)\nonumber\\
&&\left.+\frac{1}{3}b(q')b(q+q')\frac{1}{(q+q')^2}
\left(\frac{(qq')^2}{{q'}^2}
-q^2\right)\right]\nonumber\\
&&-\frac{2}{3}P^{-1}_{\rm gh}(q')P^{-1}_{\rm gh}(q+q')
\left[3\frac{(qq')}{q^2}+4\frac{(qq')^2}{(q^2)^2}-
\frac{{q'}^2}{q^2}\right]\Biggr\rbrace\ear
where
\be\label{F.68}
b(q)=\frac{(H_A(q)+\tilde R_k(q))q^2}{G_A(q)+R_k(q)+(H_A(q)
+\tilde R_k(q))q^2}\ee
For $\alpha\to0$ we observe that $H_A$ diverges
$\sim\frac{1}
{\alpha}$ and $b(q)$ approaches one. Defining as above a renormalized
$\alpha_R$ one sees that $\alpha_R=0$ is now an infrared stable fixed
point\footnote{We choose $\tilde R_kq^2=\left(\frac{1}{\alpha_R}-1
\right)R_k$ such that $b(q)=1+0(\alpha_R)$ for all values of $q^2$.}.
We further observe that for $\alpha_R\to0$ the function $F_2$ defined
in eq. (\ref{5.8}) equals $F_1/(p_1-p_3)^2$ and only the function
$G_A(q)$ determines the effective heavy four-quark interaction.
The flow
equation for $G_A$ can be written in a more compact form using
\be\label{F.68a}
P_A(q)=\tilde Z^{-1}_F(G_A(q)+R_k(q))\ee
and we approximate the ghost part of $\tilde \Gamma_k^{(2)}$ by
\be\label{F.69}
P_{\rm gh}(q)=P_A(q),\qquad \tilde\partial_tP_{\rm gh}(q)
= \tilde\partial_tP_A(q).\ee
This yields the flow euqation for the propagator for $\alpha_R=0$
\bear\label{F.70}
&&\frac{\partial}{\partial t}G_A(q)=N_cg^2_k\tilde Z_F\int\frac{d^4q'}
{(2\pi)^4} \tilde\partial_t\Biggl\lbrace
\frac{9}{4}P_A^{-1}(q')\nonumber\\
&&-\frac{1}{6}P^{-1}_A(q')P^{-1}_A(q+q')\Biggl[13q^2-14(qq')+10{q'}^2
\nonumber\\
&&-10\frac{(qq')^2}{q^2}-22\frac{(qq')^2}{{q'}^2}-4\frac{(qq')^3}
{q^2{q'}^2}+\frac{q^2}{{q'}^2}\frac{q^2{q'}^2-(qq')^2}
{(q+q')^2}\Biggr]\Biggr\rbrace\ear

The r.h.s. of the
flow equation (\ref{F.70}) involves not only $G_A(q)$ but also
the renormalized gauge coupling
$g_k$ and the gluon wave function renormalization constant $\tilde Z_F$.
We  approximate $\tilde Z_F$ by the coefficient
of the $q^2$
term in $G_A$. More precisely, we expand for small $q^2$,
$G_A(q)=\bar m^2_A+G^{(1)}_Aq^2+G_A^{(2)}(q^2)^2+...$,
and identify $\tilde Z_F$ with $G_A^{(1)}$.  For the
flow equation for the renormalized gauge coupling $g_k$ we rely
on the fact that the first two coefficients of the $\beta$ function
\bear\label{F.71}
&&\frac{\partial g^2_k}{\partial t}=\beta_{g^2}=-c_1\frac{g^4_k}
{16\pi^2}-c_2\frac{g^6_k}{(16\pi^2)^2}-...\nonumber\\
&&c_1=\frac{22N_c}{3}\qquad c_2=\frac{204}{9}N^2_c\ear
are universal
In the region of large $g_k$ we may also use nonperturbative estimates
of $\beta_{g^2}$ derived by different methods \cite{Reu}. An ansatz for
$\beta_{g^2}$ combined with an estimate of $\tilde\eta_F$ fixes also
the evolution of $\tilde g^2$ and provides all information needed for
a numerical solution of the flow equation (\ref{F.70}).

\end{document}